\documentclass[aps,prd,twocolumn,superscriptaddress,nofootinbib]{revtex4-1}

\usepackage{latexsym}
\usepackage{amsmath}
\usepackage{amssymb}
\usepackage{amsfonts}
\usepackage{bm}
\usepackage{physics}

\usepackage{color}
\definecolor{purple}{rgb}{0.5,0,0.5}
\definecolor{blue}{rgb}{0.0,0,0.9}
\definecolor{prdblue}{rgb}{0.133,0.118,0.498}
\usepackage[colorlinks=true, pdfstartview=FitV, linkcolor=prdblue, citecolor= prdblue, urlcolor=prdblue]{hyperref}

\usepackage{supertabular} 
\usepackage{placeins}
\usepackage{epsfig}
\usepackage{graphicx}

\usepackage{soul} 
\usepackage{color}

\def\tstrut{\vrule height3.25ex depth0pt width0pt} 

\usepackage{lipsum}
\usepackage{amsthm}
\usepackage{bbold}
\usepackage{dcolumn}
\usepackage{longtable}
\usepackage{epstopdf}
\usepackage{xspace}
\usepackage{cancel}
\usepackage{nicefrac}
\usepackage{multirow}
\usepackage{float}

\begin{document}

\title{Insights into the $\mathbf{\gamma^{(*)} + N(940)\frac{1}{2}^+ \to \Delta(1700)\frac{3}{2}^{-}}$ transition}

\author{L. Albino}
\email[]{luis.albino.fernandez@gmail.com}
\affiliation{Departamento Sistemas F\'isicos Qu\'imicos y Naturales, Universidad Pablo de Olavide, Sevilla, 3800708, Spain.}
\affiliation{Departmento de Ciencias Integradas, Universidad de Huelva, E-21071 Huelva, Spain.}

\author{G. Paredes-Torres}
\email[]{gustavo.paredes@umich.mx}
\affiliation{Instituto de F\'{i}sica y Matem\'aticas, Universidad Michoacana de San Nicol\'as de Hidalgo, Morelia, Michoac\'an
58040, M\'{e}xico.}
\affiliation{The Abdus Salam ICTP, Strada Costiera 11, 34151 Trieste, Italy.}
\affiliation{SISSA, via Bonomea 265, 34136, Trieste, Italy.}

\author{K. Raya}
\email[]{khepani.raya@dci.uhu.es}
\affiliation{Departmento de Ciencias Integradas, Universidad de Huelva, E-21071 Huelva, Spain.}

\author{A. Bashir}
\email[]{adnan.bashir@umich.mx}
\affiliation{Departmento de Ciencias Integradas, Universidad de Huelva, E-21071 Huelva, Spain.}
\affiliation{Instituto de F\'{i}sica y Matem\'aticas, Universidad Michoacana de San Nicol\'as de Hidalgo, Morelia, Michoac\'an
58040, M\'{e}xico.}

\author{J. Segovia}
\email[]{jsegovia@upo.es}
\affiliation{Departamento Sistemas F\'isicos Qu\'imicos y Naturales, Universidad Pablo de Olavide, Sevilla, 3800708, Spain.}

\date{\today}

\begin{abstract}
We report novel theoretical results for the $\gamma^{(*)} + N(940)\frac{1}{2}^+ \to \Delta(1700)\frac{3}{2}^{-}$ transition, utilizing a symmetry-preserving treatment of a vector$\,\otimes\,$vector contact interaction (SCI) within the Dyson-Schwinger equations (DSEs) formalism.
In this approach, both nucleon, $N(940)\frac{1}{2}^+$, and $\Delta(1232)$'s parity partner, $\Delta(1700)\frac{3}{2}^{-}$, are treated as quark-diquark composites, with their internal structures governed accordingly by a tractable truncation of the Poincaré-covariant Faddeev equation. 
Nonpointlike quark+quark (diquark) correlations within baryons, which are deeply tied to the processes driving hadron mass generation, are inherently dynamic in the sense that they continually break apart and recombine guided by the Faddeev kernel. 
For the nucleon, isoscalar-scalar and isovector-axial-vector diquarks dominate, while the $\Delta(1700)\frac{3}{2}^{-}$ state only includes contributions from isovector-axial-vector diquarks because the SCI-interaction excludes isovector-vector diquarks. 
Once the Faddeev wave function of the baryons involved in the electromagnetic transition is normalized taking into account that its elastic electric form factor must be one at the on-shell photon point, we compute the transition form factors that describe the $\gamma^{(*)} + N(940)\frac{1}{2}^+ \to \Delta(1700)\frac{3}{2}^{-}$ reaction and, using algebraic relations, derive the corresponding helicity amplitudes. 
When comparing with experiment, our findings highlight a strong sensitivity of these observables to the internal structure of baryons, offering valuable insights. Although the SCI-framework has obvious limitations, its algebraic simplicity provides analytical predictions that serve as useful benchmarks for guiding more refined studies within QCD-based DSEs frameworks.
\end{abstract}


\maketitle


\section{Introduction}
\label{sec:Intro}

Quantum Chromodynamics (QCD) is universally accepted as the fundamental theory of strong interactions, governing the dynamics of quarks and gluons via their strong interactions (see, for instance, the survey in the Review of Particle Physics of Particle Data Group~\cite{ParticleDataGroup:2024cfk}). The QCD Lagrangian, written in terms of six quark flavors with three colors each in the fundamental representation and eight colored gluons in the adjoint representation, obeys important symmetries, the most significant of which is the local color gauge invariance. Despite the elegance and apparent simplicity of the QCD Lagrangian, the conventional hadron spectrum, including baryon and meson masses, does not appear straightforwardly. Instead, it emerges from non-perturbative phenomena, such as color confinement~\cite{Alkofer:2006fu}, dynamical chiral symmetry breaking (DCSB)~\cite{Mitter:2014wpa}, and the generation of an effective gluon mass scale, see for example Ref.~\cite{Papavassiliou:2022wrb}. These mechanisms are responsible for providing mass to nucleons, and other hadrons, which can barely be accounted for solely by the Higgs mechanism~\cite{Englert:2014zpa, Higgs:2014aqa}. It only generates a minuscule proportion of the light quark mass, \emph{i.e.} the current quark masses.

Concerning the proton, it is the core of the hydrogen atom, lies at the heart of every nucleus, and has never been observed to decay. However, it is a composite object, with the valence quark content: two up ($u$) quarks and one down ($d$) quark. Focusing either solely on the proton, or the so-called nucleon when combined with its isospin partner the neutron, does not suffice to fully explain QCD's non-perturbative regime. Its excited states, named collectively $N^\ast$ nucleon resonances, offer a deeper understanding of QCD's internal dynamics and they are crucial for exploring how QCD builds the baryon spectrum. It is in this context that high-luminosity experimental facilities such as the Thomas Jefferson National Accelerator Facility (JLab) in USA~\cite{Dudek:2012vr, Accardi:2023chb}, MAMI and ELSA in Germany~\cite{Tiator:2011pw, Tiator:2018pjq} or J-PARC in Japan~\cite{Aoki:2021cqa} have been designed in order to measure the electromagnetic excitation of nucleon resonances, $\gamma^{(*)} + N \rightarrow N^\ast$, to uncover the baryon spectrum and, at the same time, to extract their transition electro-couplings, $g_vNN^\ast$, from meson electro-production data~\cite{Crede:2013kia}. Obtained primarily with the CLAS detector at the JLab, the electro-couplings of all low-lying $N^\ast$ states with mass less than $1.6\,\text{GeV}$ have been determined via independent analyses of $\pi^+ n$, $\pi^0 p$ and $\pi^+ \pi^- p$ exclusive channels; and preliminary results for the $g_v N N^\ast$ electro-couplings of most higher-lying $N^\ast$ states with masses below $1.8\,\text{GeV}$ have also been obtained from CLAS meson electro-production data~\cite{CLAS:2009ces, Aznauryan:2012ba, Proceedings:2020fyd}.

A coupled formalism of DSEs, Bethe-Salpeter equation and Faddeev equation provides, at least in principle, an ideal theoretical framework to compute transition electro-couplings, $g_vNN^\ast$. This framework naturally links both the infrared and ultraviolet behavior of the hadronic observables as it makes no recourse to the strength of the strong interaction. Moreover, it connects the QCD's fundamental degrees of freedom -- quarks and gluons -- with the observable properties of mesons and baryons~\cite{Maris:2003vk, Roberts:1994dr, Yin:2021uom}; in particular, their elastic and transition electromagnetic form factors~\cite{Eichmann:2011vu, Sanchis-Alepuz:2017mir}. However, the fact remains that all practical attempts to extract physically acceptable and trust-able solutions requires a great deal of systematic effort and continuous improvement. A remarkable progress over the years has resulted in this formalism contributing successfully even to the precision observables of the standard model of particle physics, see, for example, Refs.~\cite{Raya:2019dnh, Miramontes:2021exi, Miramontes:2024fgo, Aoyama:2020ynm}. 

One of the key insights from DSEs studies that employ realistic quark-quark interactions~\cite{Qin:2011dd, Binosi:2014aea} is the emergence of nonpointlike quark$+$quark (diquark) correlations within baryons~\cite{Maris:2002yu, Eichmann:2008ef, Cloet:2011qu, Eichmann:2016yit, Barabanov:2020jvn}. Empirical evidence in support of the presence of diquarks in the proton is accumulating~\cite{Close:1988br, Cloet:2005pp, Cates:2011pz, Segovia:2015ufa, Cloet:2012cy, Cloet:2014rja}. It is worth reiterating that these correlations are not the elementary diquarks introduced roughly fifty years ago in order to simplify treatment of the three-quark bound-state problem~\cite{Lichtenberg:1967zz, Lichtenberg:1968zz}. The two-body correlation predicted by modern pseudo-Faddeev equation studies is dynamic; the dressed-quarks participate in all diquark clusters. Noticeably, the baryon spectrum produced through the quark-diquark picture has significant overlap with that predicted from the three dressed-quark scenario, lattice-QCD numerical calculations and, of course, experiment (see Refs.~\cite{Eichmann:2016yit, Barabanov:2020jvn} for reviews).

The quark-diquark picture of baryons, combined with symmetry-preserving quark-quark interaction kernels and vertices that possess either QCD-like momentum dependence or simpler contact terms, has proven particularly successful in describing the transition electro-couplings of low-lying nucleon resonances such as ground states: $N(940)\frac{1}{2}^{+}$~\cite{Wilson:2011aa, Cloet:2008re, Segovia:2014aza, Cui:2020rmu, Yao:2024uej} and $\Delta(1232)\frac{3}{2}^{+}$~\cite{Segovia:2013rca, Segovia:2013uga, Segovia:2014aza, Segovia:2016zyc}; first radial excitations: $N(1440)\frac{1}{2}^{+}$~\cite{Wilson:2011aa, Segovia:2015hra, Chen:2018nsg} and $\Delta(1600)\frac{3}{2}^{+}$~\cite{Lu:2019bjs}; and the parity partner of the nucleon $N(1535)\frac{1}{2}^{-}$~\cite{Raya:2021pyr}. What has so far been missing is the theoretical calculation of electromagnetic (transition) form factors for the case $\Delta(1700)\frac{3}{2}^{-}$. This is the very object of study in this manuscript. We compute the transition electromagnetic form factors for the $\gamma^{(*)} + N(940)\frac{1}{2}^+\to \Delta(1700)\frac{3}{2}^{-}$ process. Then, using appropriate linear algebraic relations, we derive the corresponding helicity amplitudes from the explicit expressions of the form factors. All these can be compared with experimental measurements at various photon virtualities~\cite{ParticleDataGroup:2024cfk, Burkert:2002zz, CLAS:2009tyz, Mokeev:2013kka, Mokeev:2020hhu, Mokeev:2024beb}, enabling insights into the internal structure of the $\Delta(1700)\frac{3}{2}^{-}$ resonance. Within the nucleon, $N(940)\frac{1}{2}^+$, the isoscalar-scalar and isovector-axial-vector diquarks play dominant roles, while for the $\Delta(1700)\frac{3}{2}^{-}$, only the isovector-axial-vector diquark contribute. The exclusion of isovector-vector diquark correlations is a direct consequence of using a symmetry-preserving contact interaction (SCI) within the DSEs approach adopted in this work.

Moreover, it is in line with the results already made available by employing more sophisticated DSEs frameworks~\cite{Liu:2022ndb}. Though, this model oversimplifies the dynamics of relativistic bound-state systems, it still retains crucial QCD features such as gauge invariance, quark confinement and dynamical chiral symmetry breaking. Our findings highlight a strong sensitivity of these observables to the internal structure of baryons, offering valuable insight. Although the SCI-framework has obvious limitations, its algebraic simplicity provides analytical predictions that serve as useful benchmarks for guiding more refined studies within QCD-based DSEs frameworks.

After this comprehensive introduction, the manuscript is organized as follows. Section~\ref{sec:Theory} is devoted to present the fundamental aspects of our theoretical framework, beginning with a brief description of the symmetry-preserving vector$\,\otimes\,$vector contact interaction (SCI) and the key model parameters, see Subsection~\ref{subsec:SCI}. We introduce the Faddeev amplitudes for both the nucleon and $\Delta$ baryons in Subsection~\ref{subsec:Faddeev}, emphasizing the role of diquarks and their importance in dictating the structure of these baryons. Subsection~\ref{subsec:EFFs} provides a detailed description of the electromagnetic interactions which contribute to the elastic and transition form factors in our theoretical framework. In Subsection~\ref{subsec:Currents}, we present the general decomposition of the electromagnetic current for the nucleon, delta and nucleon-to-delta expressed in terms of electromagnetic form factors. We also provide the relations between helicity amplitudes and transition form factors. In Sec~\ref{sec:Results}, we discuss our numerical results for both transition form factors and helicity amplitudes, comparing them with experimental measurements. Finally, we provide a brief summary of our findings and an outlook for future in Sec~\ref{sec:Summary}.


\section{THEORETICAL FRAMEWORK}
\label{sec:Theory}

\subsection{Symmetry-preserving contact interaction (SCI)}
\label{subsec:SCI}

\begin{figure}[!t]
\centerline{%
\includegraphics[clip,width=0.45\textwidth, height=0.08\textheight]{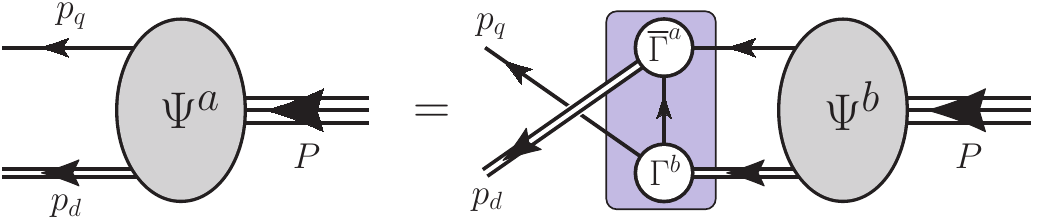}}
\caption{\label{fig:Faddeev} Poincar\'e covariant Faddeev equation.  $\Psi$ is the Faddeev amplitude for a baryon of total momentum $P= p_q + p_d$.  The shaded rectangle demarcates the kernel of the Faddeev equation: \emph{single line}, dressed-quark propagator; $\Gamma$,  diquark correlation amplitude; and \emph{double line}, diquark propagator.}
\end{figure}

We base our description of baryon bound-states on a Poincar\'e-covariant Faddeev equation, which is illustrated in Fig.~\ref{fig:Faddeev}. Its key elements are the dressed-quark and -diquark propagators, and the diquark Bethe-Salpeter amplitudes. All are completely determined once the quark-quark interaction kernel is specified and, as explained in the Introduction, we use a symmetry-preserving regularization of a vector$\,\otimes\,$vector contact interaction, which is characterized by a constant gluon propagator,
\begin{eqnarray}
g^2 D_{\mu\nu} \left( p-q \right) = \frac{4 \pi \alpha_{\text{IR}}}{m_g^2} \delta_{\mu\nu} \,,
\label{eq:GluonPropagator}
\end{eqnarray}
where the effective gluon mass $m_{g} = 0.5$ GeV is consonant with the mass-scale characteristic of a non-zero, IR finite gluon propagator in lattice regularized QCD~\cite{Bowman:2004jm, Bogolubsky:2009dc, Aguilar:2008xm}, and 
$\alpha_{IR} = 0.36\pi$ plays the role of an effective coupling whose fitted value is commensurate with contemporary estimates of the zero-momentum value of a running-coupling in QCD~\cite{Binosi:2016nme, Cui:2019dwv}.

We embed Eq.~\eqref{eq:GluonPropagator} in a rainbow-ladder (RL) truncation of the DSEs.\footnote{The RL truncation entails a tree-level quark-gluon vertex, namely $\Gamma_{\mu}^a \left( q,p \right) = \frac{\lambda^a}{2} \gamma_{\mu}$, where $\lambda^a$ stand for the Gell-Mann matrices.} Therefore, it yields the following dressed-quark propagator $S(p)$:
\begin{eqnarray}
\label{Dressed-quark propagator}
S^{-1}(p) = i \gamma \cdot p + M \,, 
\end{eqnarray}
with the momentum-independent dressed-quark mass, $M$, determined as
\begin{eqnarray}
\label{non-regularized gap equation}
M = m_0 + M \frac{4 \alpha_{IR}}{3 \pi m_g^2} \int_{0}^{\infty}{ds \frac{s}{s + M^2}} \,, 
\end{eqnarray}
where $m_0$ is the current-quark mass. Equation~\eqref{non-regularized gap equation} possesses a quadratic divergence. Therefore, it needs to be regularized in a Poincar\'e covariant manner. The procedure involves a proper-time regularization~\cite{Ebert:1996vx} which entails an infrarred cut-off $\tau_{ir}$ that implements confinement~\cite{Krein:1990sf, Chang:2011vu, Roberts:2012sv} and an ultraviolet cut-off $\tau_{uv}$ that sets the scale of dimensionless quantities~\cite{Gutierrez-Guerrero:2010waf}:
\begin{eqnarray}
\frac{1}{s + M^2} \to \frac{\mathrm{e}^{- \left( s + M^2 \right) \tau_{uv}^2} - \mathrm{e}^{- \left( s + M^2 \right) \tau_{ir}^2}}{s + M^2} \,. \label{proper-time regularization}
\end{eqnarray}
Using the above regularized denominator, Eq.~\eqref{proper-time regularization}, in the dressed-quark gap equation, Eq.~\eqref{non-regularized gap equation}, yields an integral equation for $M$ whose solution entails a non-zero dressed-quark mass even in the chiral limit ($m_0=0$). In this work, we employ the set of model parameters implemented in Ref.~\cite{Yin:2021uom}, namely $1/\tau_{ir} = 0.240$ GeV and $1/\tau_{uv} = 0.905$ GeV, adjusted to fit both the pion mass and its decay constant. Such a set of parameters results in a constituent quark mass $M=0.37$ GeV for both up and down quarks (assuming isospin symmetry) provided $m_0 = 7$ MeV.

\subsection{The $\mathbf{N}$ and $\mathbf{\Delta}$ Faddeev equations}
\label{subsec:Faddeev}

Within our quark-diquark picture, the baryon's wave function is compactly expressed as
\begin{eqnarray}
\Psi = \Psi^1 +\Psi^2 +\Psi^3 \,,
\end{eqnarray}
where the superscript stands for the spectator quark, \emph{e.g.} $\Psi^{1,2}$ are obtained from $\Psi^3$ by a cyclic permutation of all the quark labels. As shown in Refs.~\cite{Chen:2017pse, Chen:2019fzn, Liu:2022ndb}, using a framework built upon a Faddeev equation kernel and interaction vertices that possess QCD-like momentum dependence, the $N(940)\frac{1}{2}^+$ and $\Delta(1700)\frac{3}{2}^-$ baryons are heavily dominated by isoscalar-scalar and isovector-axial-vector diquark correlations. Therefore, we assume that a simple but realistic representation of the Faddeev amplitude of a positively charged nucleon is given by
\begin{eqnarray}
\label{Nucleon FA}
\Psi^3 &=& \Gamma^{0^{+}}(p_1,p_2)\Delta^{0^{+}}(K)\mathcal{S}(P)u(P) \nonumber \\
&& \hspace{-.5cm} +\sum_{j=1,2} \Gamma^{1^{+}_j}_{\alpha}(p_1,p_2)\Delta^{1^{+}}_{\alpha\beta}(K)\mathcal{A}^j_{\beta}(P)u(P) \,,
\end{eqnarray}
and, for the $\Delta^{+}$-baryon, we have
\begin{eqnarray}
 \label{Delta FA}
\Psi^3 = \sum_{j=1,2} \Gamma^{1^{+}_j}_{\alpha}(p_1,p_2)\Delta^{1^{+}}_{\alpha\beta}(K)\mathcal{D}^{j}_{\beta\rho}(P)u_{\rho}(P) \,.
\end{eqnarray}
The spinor $u(P)$ satisfies the Dirac equation 
\begin{equation}
\left(i\gamma \cdot P + m_B\right)u(P) = 0 \,,
\end{equation}
for on-shell nucleons with momentum $P$, and $u_{\rho}(P)$ is the Rarita-Schwinger spinor representing on-shell Delta baryons.

In the above amplitudes, the flavor structure has been omitted for the sake of simplified notation, but the reader can refer to Ref.~\cite{Chen:2012qr} for a more comprehensive overview. Meanwhile, $\Gamma^{0^{+}}(p_1,p_2)$ stands for the canonically-normalized Bethe-Salpeter amplitude of isoscalar-scalar diquark $0^{+}=\left[ ud \right]$. The $\Gamma^{1^{+}_j}(p_1,p_2)$, with $j=1,2$, is the corresponding canonically-normalized Bethe-Salpeter amplitude of isovector-axial-vector diquark with flavor content $1^{+}_1=\left\{ uu \right\}$ and $1^{+}_2=\left\{ ud \right\}$. Both diquark amplitudes depend on constituent quark momenta $p_1$ and $p_2$, resulting in a diquark's total momentum $K=p_1+p_2$. In addition, scalar and axial-vector diquark propagators are respectively expressed as
\begin{eqnarray}
\Delta^{0^{+}} (K) &=& \frac{1}{K^2 + m^2_{0^{+}}} \,, \label{Diquark-propagator: scalar} \\
\Delta^{1^{+}}_{\mu \nu} (K)  &=& \left[ \delta_{\mu \nu} + \frac{K_{\mu} K_{\nu}}{m^2_{1^{+}}} \right] \frac{1}{K^2 + m^2_{1^{+}}} \,, \label{Diquark-propagator: axial}
\end{eqnarray}
where $m_{0^{+}}$ and $m_{1^{+}}$ are the mass-scales associated with the corresponding diquark correlation. In this work, we use the values obtained in Ref.~\cite{Yin:2021uom}:
\begin{eqnarray}
m_{\left[ ud \right]_{0^+}} = 0.78 \, \hbox{GeV} \,, \\
m_{\left\{ uu \right\}_{1^+}} = m_{\left\{ ud \right\}_{1^+}} = m_{\left\{ dd \right\}_{1^+}} =  1.06 \, \hbox{GeV} \,.
\end{eqnarray}

The general form of the matrices $\mathcal{S}(P)$, $\mathcal{A}^j_{\beta}(P)$ and $\mathcal{D}^{j}_{\beta\rho}(P)$ describing the quark-diquark momentum correlations in Nucleon and Delta baryons, Eqs.~\eqref{Nucleon FA} and~\eqref{Delta FA}, are simplified in the SCI to
\begin{eqnarray}
\mathcal{S}(P) &=& s \, I_D \,,  \label{Faddeev Amplitude - Nucleon: scalar part} \\
\mathcal{A}^j_{\beta}(P) &=& i \, a^{1^{+}_j}_1 \gamma_5 \gamma_{\beta} + a^{1^{+}_j}_2 \gamma_5 \hat{P}_{\beta} \,, \label{Faddeev Amplitude - Nucleon: axial-vector part} \\
\mathcal{D}^{j}_{\beta\rho}(P) &=& d^{1^{+}_j} \, \delta_{\beta\rho} \,, \label{Faddeev Amplitude - Delta: scalar part}
\end{eqnarray}
with $\mathbf{I}_{\mathrm{D}}$ defining the $4 \times 4$ identity matrix in Dirac space and $a^{\left\{ ud \right\}}_i = - a^{\left\{ uu \right\}}_i/\sqrt{2}$, for both $i=1,2$, as well as $d^{\left\{ ud \right\}}=\sqrt{2} d^{\left\{ uu \right\}}$. 

As apparent in Fig.~\ref{fig:Faddeev}, the Faddeev kernel (shaded rectangle) involves diquark breakup and reformation via exchange of a dressed-quark. We now follow Ref.~\cite{Roberts:2011cf} and make a marked but convenient simplification with impunity~\cite{Xu:2015kta}; namely, in the Faddeev equation for a baryon of type $B$, the quark exchanged between the diquarks is represented as
\begin{equation}
S^T=\frac{g_B^2}{M} \,,
\end{equation}
where the superscript $T$ indicates matrix transpose, $M$ is the dressed quark mass and $g_B$ is an effective coupling whose numerical values are $g_N=1.18$ and $g_\Delta=1.56\,g_{+-}$ for the $N(940)\frac{1}{2}^+$ and $\Delta(1700)\frac{3}{2}^-$ baryons, respectively; $g_{+-}=\sqrt{0.1}$ is a factor which takes into account the difference in parity between the diquark and the host baryon~\cite{Yin:2019bxe, Yin:2021uom}. This is a variant of the so-called \emph{static approximation}, which itself was introduced in Ref.~\cite{Buck:1992wz} and has subsequently been used in studies of a wide range of baryon properties~\cite{Yin:2019bxe, Yin:2021uom}.

All together, the resulting Faddeev equations yield
\begin{eqnarray}
m_{N(940)} &=& 1.14 \, \hbox{GeV} \,, \nonumber \label{Nucleon 940 mass}\\
m_{\Delta(1700)} &=& 1.72 \, \hbox{GeV} \,, \nonumber \label{Delta 1700 mass}
\end{eqnarray}
for the baryon masses and 
\begin{center}
\begin{tabular}[t]{c|c c c|c}
& $s$ & $a^{\left\{ uu \right\}}_1$ & $a^{\left\{ uu \right\}}_2$ & \; $d^{\left\{ uu \right\}}$ \; \\
\hline
\tstrut
$N(940)$ & \; 0.88 \; & \; -0.38 \; & \; -0.06 \; & \; - \; \\[1ex]
$\Delta(1700)$ & \; - \; & \; - \; & \; - \; & \; 0.58 \; \\
\end{tabular}
\end{center}
for the Faddeev amplitudes.


\begin{figure}[!t]
\centering
\includegraphics[scale=0.60]{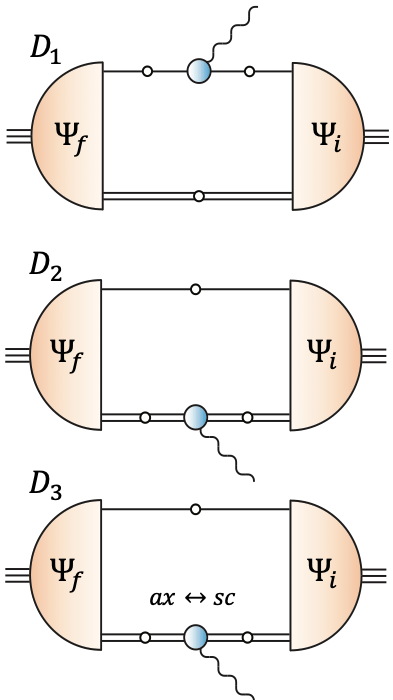}
\caption{\label{fig:EMInteractions} Diagrammatic representation of contributions for elastic and transition EM currents in the quark-diquark picture. Dressed-quark and diquark propagators are represented by single and double lines, respectively. Initial ($\Psi_i$) and final ($\Psi_f$) state's Faddeev amplitude for the involved baryons are represented by orange semi-circles. The blue blobs represent the corresponding quark-photon and diquark-photon vertices for each of the following diagrams: in $D_1$ it entails the photon coupling to a dressed-quark; in $D_2$ the photon couples elastically to a diquark; and $D_3$ involves photon-induced transitions between axial-vector and scalar diquarks.}
\end{figure}

\subsection{The photon-baryon coupling}
\label{subsec:EFFs}

Within our theoretical framework, one needs to compute the $N(940)\frac{1}{2}^+$ and $\Delta(1700)\frac{3}{2}^-$ elastic form factors primarily because the $Q^2 = 0$ value of the leading electric form factor is required in order to normalize the Faddeev amplitudes. It is necessary if one wishes to analyze later the $\gamma^{(\ast)} + N(940)\frac{1}{2}^+ \to \Delta(1700)\frac{3}{2}^-$ transition, which is empirically accessible~\cite{ParticleDataGroup:2024cfk, Burkert:2002zz, CLAS:2009tyz, Mokeev:2013kka, Mokeev:2020hhu, Mokeev:2024beb}.

Therefore, the elastic and transition form factors of interest can be derived from the following current: 
\begin{eqnarray}
\mathcal{J}^{B_f B_i}_{\mu \left[ \rho \right] \left[ \sigma \right]}(P_f,P_i) && \nonumber \\
&& \hspace{-2cm} = \hspace{-.2cm} \sum_{n=1,2,3} \int_{dk}{ \mathcal{P}^{B_f}_{ \left[ \rho \alpha \right]} \mathcal{G}_f^{\pm} \mathcal{G}_{f/i} \Lambda^{D_n}_{\mu \left[ \alpha \right] \left[ \beta \right]} \left(k; P_f, P_i\right) \mathcal{G}_i^{\pm} \mathcal{P}^{B_i}_{ \left[ \beta \sigma \right]} } \,, \nonumber \\
\end{eqnarray}
where $B_{i(f)}$ denotes the initial and final baryon states, with incoming and outgoing momenta $P_i$ and $P_f$, respectively. Moreover, the integration notation entails $\int_{dk} = \int d^4k/(2\pi)^4$ and the sum is performed over the possible quark-photon and diquark-photon contributions $\Lambda^{D_n}_{\mu \left[ \alpha \right] \left[ \beta \right]} \left(k; P_f, P_i\right)$ allowed in our quark-diquark picture, diagrammatically represented in Fig.~\ref{fig:EMInteractions}, and described in the subsequent subsections. Furthermore, baryon's parity is reflected via the Dirac structures $\mathcal{G}^{+(-)}_{i,f} = \mathbf{I}_{\mathrm{D}} (\gamma_5)$, the additional $\mathcal{G}_{f/i}$ is equal to $\gamma_5$ for a flipping parity transition between initial and final baryons and it is $\mathbf{I}_{\mathrm{D}}$ for baryon's parity conservation in the transition. Besides, Lorentz indices denote photon and Delta baryon polarizations: $\mu$ is associated with the photon whereas Greek letters between square brackets indicate that the indices will only appear if they are associated with the appearance of a Delta-baryon in either the initial or final state. For nucleon, the operator $\mathcal{P}^B_{\left[ \rho \alpha \right]} (P)$ reduces to the positive-energy projector $\Lambda_{+}(P) = (1-i \gamma \cdot \hat{P})/2$, with $\hat{P} = P/m_N$ normalized to the nucleon mass. For the Delta-baryon, the operator $\mathcal{P}^B_{\left[ \rho \alpha \right]} (P)$ is again the positive energy projector, changing $m_N$ by $m_\Delta$, and extended by the Rarita-Schwinger projection operator $\mathcal{R}_{\rho \alpha} (P)$. Namely,
\begin{eqnarray}
\mathcal{P}^N (P) & = & \Lambda_{+}(P) \,, \\
\mathcal{P}^{\Delta}_{\rho \alpha} (P) & = & \Lambda_{+}(P) \mathcal{R}_{\rho \alpha} (P) \,,
\end{eqnarray}
where
\begin{eqnarray}
\mathcal{R}_{\rho \alpha} (P) & = & \delta_{\rho \alpha} -\frac{1}{3} \gamma_{\rho} \gamma_{\alpha} \nonumber \\
&& +\frac{2}{3} \hat{P}_{\rho} \hat{P}_{\alpha} +\frac{1}{3} \left( \gamma_{\rho} \hat{P}_{\alpha} - \gamma_{\alpha} \hat{P}_{\rho} \right) \,.
\end{eqnarray}

Conservation of photon momentum $Q$, along with the on-shell condition $P^2_{i(f)} = -m^2_{i(f)}$, yields to the following kinematic relations ($Q = P_f -P_i$):
\begin{eqnarray}
Q \cdot P_f &=& \frac{1}{2} \left( m_i^2 - m_f^2 + Q^2 \right) \,, \\
Q \cdot P_i &=& \frac{1}{2} \left( m_i^2 - m_f^2 - Q^2 \right) \,, \\
P_f \cdot P_i &=& -\frac{1}{2} \left( m_i^2 + m_f^2 + Q^2 \right) \,.
\end{eqnarray}

Within the SCI approach and only considering isoscalar-scalar and isovector-axial-vector diquark correlations, there are three types of contributions that ensure current conservation. These are diagrammatically represented in Fig.~\ref{fig:EMInteractions}; the corresponding mathematical content is described in the following subsections.


\subsubsection{The diagram $D_1$: quark-photon coupling}

The upper diagram in Fig.~\ref{fig:EMInteractions} represents the contribution of a photon directly coupled to a quark inside baryon with an electric charge $e_q$. The corresponding interaction vertex reads
\begin{eqnarray}
\Lambda^{D_1 (J^P)}_{\mu \left[ \alpha \right] \left[ \beta \right]} \left(k; P_f, P_i\right) &=& \nonumber \\
&& \hspace{-2.8cm} e_q\mathcal{F}^{B_f}_{\left[ \alpha \right] \left[ \delta \right]} S(-k_{-f})\Lambda_{\mu}^{q\gamma q} S(-k_{-i}) \Delta^{J^P}_{ \left[ \delta \lambda \right]}(-k) \mathcal{F}^{B_i}_{\left[ \lambda \right] \left[ \beta \right]} \,, \nonumber \\
\end{eqnarray}
where $k_{\pm i(f)} = -k \pm P_{i(f)}$. Recall that square brackets indicate the Lorentz indices that might be needed depending on the diquark quantum numbers. Namely, axial-vector diquarks require a Lorentz index whereas scalar diquarks do not, Eqs.~\eqref{Diquark-propagator: scalar} and~\eqref{Diquark-propagator: axial}. Analogously, subscripts on the corresponding Faddeev amplitudes, $\mathcal{F}^{B_{i(f)}}_{\left[ \alpha \right] \left[ \beta \right]} \equiv \mathcal{F}^{B_{i(f)}}_{\left[ \alpha \right] \left[ \beta \right]} (P_{i(f)})$, depend on the related baryon and are defined in Eqs.~\eqref{Faddeev Amplitude - Nucleon: scalar part},~\eqref{Faddeev Amplitude - Nucleon: axial-vector part} and~\eqref{Faddeev Amplitude - Delta: scalar part}. Moreover, the quark-photon vertex, $\Lambda_{\mu}^{q\gamma q} \equiv \Lambda_{\mu}^{q\gamma q}(Q)$, is defined consistently with the SCI and the RL approaches
\begin{eqnarray}
\Lambda_{\mu}^{q\gamma q}(Q) = \gamma_{\mu}^{T} P_{T}(Q^2) + \eta \, \sigma_{\mu\nu}Q_{\nu} F_{AMM} \left( Q^2 \right) \,, \label{Quark-Photon vertex}
\end{eqnarray}
with $\gamma_{\mu}^{T} = \gamma_{\mu} - Q_{\mu} \gamma \cdot Q / Q^2$ and $\sigma_{\mu\nu}=i\left[ \gamma_{\mu}, \gamma_{\nu} \right]/2$, thus ensuring current conservation ($Q_{\mu} \mathcal{J}^{B_f B_i}_{\mu \left[ \rho \right] \left[ \sigma \right]} = 0$) for this contribution. The second term in Eq.~\eqref{Quark-Photon vertex}, modulated by a dimensionless factor $\eta$, encodes the effect of the large anomalous magnetic moment (AMM) characteristic of dressed light-quarks in the presence of DCSB:
\begin{eqnarray}
F_{AMM} \left( Q^2 \right) = \frac{1}{2M} \exp\left(-\frac{Q^2}{4M^2} \right) \,. \label{Vertex Dressing: quark-photon AMM} 
\end{eqnarray}
Moreover, satisfying both vector and axial-vector Ward-Takahashi identities entails~\cite{Roberts:2011wy}
\begin{eqnarray}
\hspace{-.8cm} P_{T}^{-1}(Q^2) =&& \nonumber \\ && \hspace{-1.2cm} 1+\frac{4 \alpha_{IR}}{3 \pi m_g^2} \int_0^1{d\alpha \, \alpha \left( 1- \alpha \right) Q^2 \mathcal{C}_2 \left( \omega \left( \alpha, Q^2 \right) \right) } , \label{PT from SCI}
\end{eqnarray}
with $\omega \left( \alpha ,Q^2 \right) = M^2 + \alpha \left( 1- \alpha \right) Q^2$ and
\begin{eqnarray}
\mathcal{C}_{\beta} (x) =  \frac{\omega ^{2-\beta}}{\Gamma \left[ \beta \right]} \Gamma \left[ \beta -2, x \, \tau_{uv}^2, x \, \tau_{ir}^2 \right] \,,
\end{eqnarray}
where the gamma and generalized incomplete gamma functions are used.

\begin{figure}[!t]
\centering
\includegraphics[scale=.25]{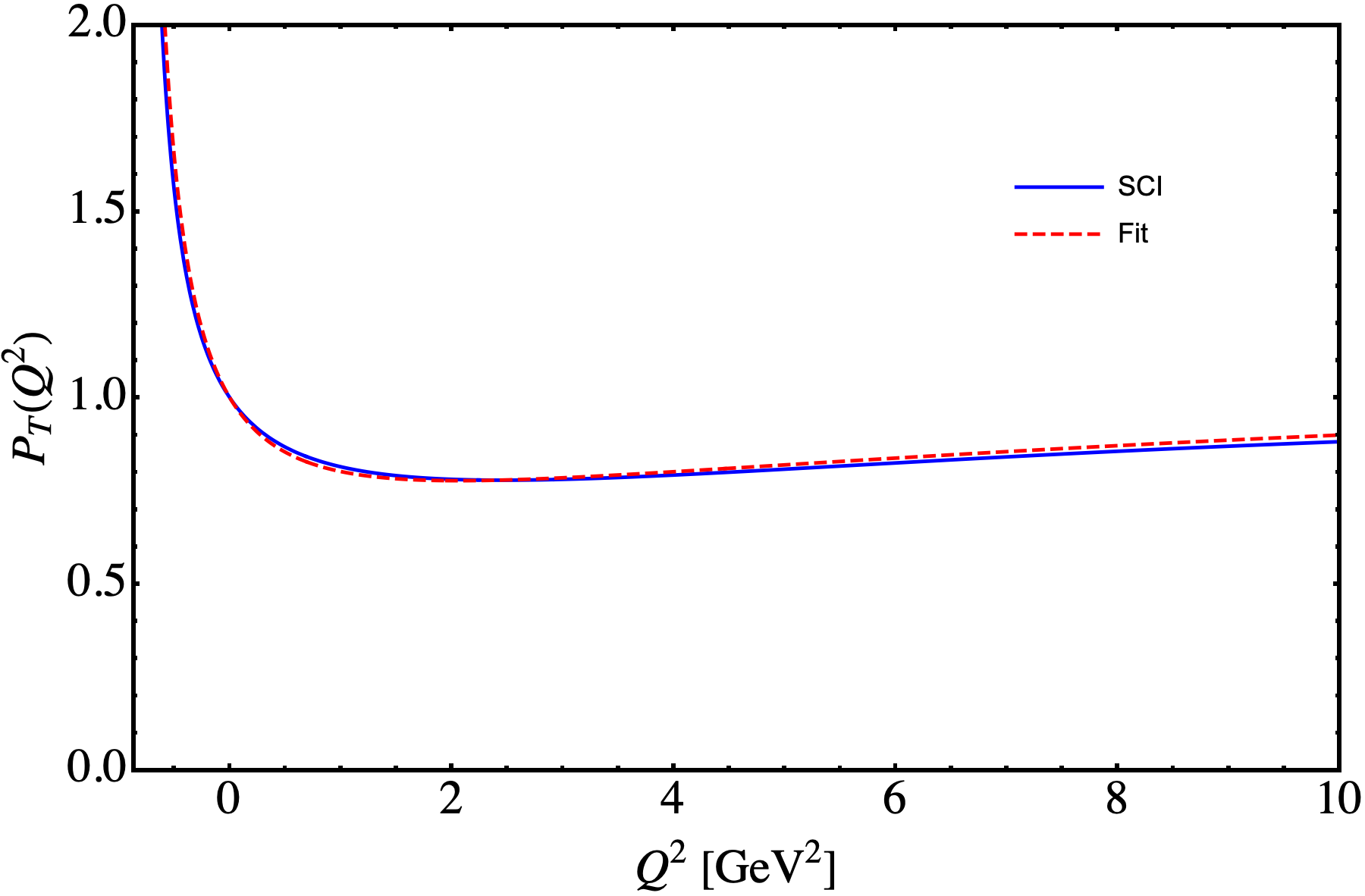}
\caption{\label{fig:PTcomparison} Quark-photon dressing function $P_T \left( Q^2 \right)$. The blue solid line is the SCI solution, Eq.~\eqref{PT from SCI}, consistent with vector and axial-vector Ward-Takahashi identities; the red dashed line is the parametization of Eq.~\ref{PT Fit}. At large spacelike-$Q^2$, the shown point-wise behavior yields $P_T \left( Q^2 \right) \rightarrow 1$, characteristic of a pointlike particle.}
\end{figure}

An amenable approximation for $P_T(Q^2)$ that parametrizes quite well the solution of Eq.~\eqref{PT from SCI} on the region $Q^2 / \hbox{GeV}^2 \in \left[ -0.75 , 10 \right]$ is shown in Fig.~\ref{fig:PTcomparison}. This fit entails a pole at $Q^2=-0.865 \, \hbox{GeV}^2$, consonant with the $\rho$-pole,  and it is parametrized as~\cite{Roberts:2011wy}
\begin{eqnarray}
P_T \left( x \right) = \frac{1+ 0.7743 \, x + 0.1548 \, x^2}{1 + 1.2706 \, x + 0.1317 \, x^2} \,. \label{PT Fit}
\end{eqnarray}


\subsubsection{The diagram $D_2$: elastic diquark-photon coupling}

The elastic scattering of a photon from a diquark with electric charge $e_{J^P}$ is pictorially represented by the middle diagram of Fig~\ref{fig:EMInteractions}. This might be characterized as follows:
\begin{eqnarray}
\Lambda^{D_2 \left( J^P \right)}_{\mu \left[ \alpha \right] \left[ \beta \right]} \left(k; P_f, P_i\right) &=& \nonumber \\
&& \hspace{-2.8cm} e_{J^P} \mathcal{F}^{B_f}_{\left[ \alpha \right] \left[ \delta \right]} S(k)\Delta^{J^P}_{ \left[ \delta \rho \right]}(k_f) \Lambda_{\mu \left[ \rho \sigma \right]}^{\gamma J^P} \Delta^{J^P}_{ \left[ \sigma \lambda \right]}(k_i) \mathcal{F}^{B_i}_{\left[ \lambda \right] \left[ \beta \right]} \,, \nonumber \\
\end{eqnarray}
where the structure of the corresponding diquark-photon vertex, $\Lambda_{\mu \left[ \rho \sigma \right]}^{\gamma J^P} \equiv \Lambda_{\mu \left[ \rho \sigma \right]}^{\gamma J^P} (k_f, k_i)$, is determined by the diquark $J^P$-type involved in the scattering process. That is to say, one has for a scalar diquark ($2K= k_f +k_i $)
\begin{eqnarray}
i\Lambda_{\mu}^{\gamma 0^{+}} (k_f, k_i) = 2 K_{\mu} F^{0^{+}} \hspace{-.15cm} \left( Q^2 \right) \,, \label{Vertex: photon-sc diquark}
\end{eqnarray}
whereas for an axial-vector diquark
\begin{eqnarray}
i \Lambda_{\mu, \rho \sigma}^{\gamma 1^{+}} = \sum_{j=1}^{3} {T^j_{\mu, \rho \sigma} \left( K,Q \right) F_j^{1^{+}} \hspace{-.15cm} \left( Q^2 \right) } \,, \label{Vertex: photon-ax diquark}
\end{eqnarray}
with
\begin{eqnarray}
T^1_{\mu, \rho \sigma} \left( K,Q \right) &=& 2 K_{\mu} \mathcal{T}^{\,i}_{\rho \nu} \mathcal{T}^f_{\nu \sigma} \,, \\
T^2_{\mu, \rho \sigma} \left( K,Q \right) &=& Q_{\nu} \mathcal{T}_{\rho \nu} \left( k_i, -Q/2 \right) \mathcal{T}^f_{\mu \sigma} \nonumber \\
&& 
\hspace*{-0.50cm} - Q_{\nu} \mathcal{T}_{\sigma \nu} \left( k_f, Q/2 \right) \mathcal{T}^i_{\mu \rho} \,, \\
T^3_{\mu, \rho \sigma} \left( K,Q \right) &=& \frac{K_{\mu} Q_{\nu} Q_{\lambda}}{m_{1^{+}}^2} \nonumber \\
&&
\hspace*{-0.50cm} \times \mathcal{T}_{\rho \nu} \left( k_i, -Q/2 \right) \mathcal{T}_{\lambda \nu} \left( k_f, Q/2 \right) \,,
\end{eqnarray}
where $\mathcal{T}_{\rho \sigma} \left( k, p \right) = \delta_{\rho \sigma} + k_{\rho} p_{\sigma} / m_{1^{+}}^2$ and $\mathcal{T}^{i(f)}_{\rho \sigma}  = \mathcal{T}_{\rho \sigma} \left( k_{i(f)}, k_{i(f)} \right)$. 

The dressing functions in Eqs.~\eqref{Vertex: photon-sc diquark} and \eqref{Vertex: photon-ax diquark} can be split into a $\rho$-pole (RP) and an AMM contribution,  
\begin{eqnarray}
F^{J^P}_{(j)} \hspace{-.15cm} \left( Q^2 \right) = F_{RP} \hspace{-.1cm} \left( Q^2 \right) + \eta \, \tilde{F}_{AMM} \hspace{-.1cm} \left( Q^2 \right) \,, \label{Vertex Dressings decomposition}
\end{eqnarray}
such that
\begin{eqnarray}
\hspace{-.3cm} F_{RP} \left( x \right) \hspace{-.1cm} &=& \hspace{-.1cm} \frac{a_0 + a_1 x + a_2 x^2}{1 +b_1 x + b_2 x^2} \,, \label{Vertex Dressing: RP-contribution} \\
\hspace{-.3cm} \tilde{F}_{AMM} \left( x \right) \hspace{-.1cm} &=& \hspace{-.1cm} \frac{a'_0 + a'_1 x}{1 +b'_1 x + b'_2 x^2}  \exp \left( -\upsilon x \right) \,, \label{Vertex Dressing: AMM-contribution}
\end{eqnarray}
with the associated fit parameters listed in Table~\ref{Vertex Dressing Funtion Parameters}. These were already fitted while studying the transition electromagnetic form factors of the nucleon's first excited and parity partner baryons: $N(1440)\frac{1}{2}^+$~\cite{Wilson:2011aa} and $N(1535)\frac{1}{2}^-$~\cite{Raya:2021pyr}.

\begin{table*}[!t]
\centering
\begin{tabular}[t]{c|c c c c|c c c c|c}
\hline\hline
\tstrut
& \; $a_0$ \; & \; $a_1$ \; & \; $b_1$ \; & \; $b_2$ \; & \; $a'_0$ \; & \; $a'_1$ \; & \; $b'_1$ \; & \; $b'_2$ \; & \;  $\upsilon$ \; \\
\hline
\tstrut
\; $F^{0^{+}}$ \; & \; 1 \; & \; 0.209 \; & \; 1.207 \; & \; $2.35 \times 10^{-7}$ \; & \; $1.67 \times 10^{-5}$ \; & \; -0.713 \; & \; 0.633 \; & \; 0.634 \; & \; 1.714 \; \\
\; $F^{1^{+}}_1$ \; & \; 1 \; & \; 0.454 \; & \; 1.719 \; & \; 0.670 \; & \; -0.001 \; & \; -0.902 \; & \; 1.337 \; & \; 0.705 \; & \; 1.263 \; \\
\; $F^{1^{+}}_2$ \; & \; -2.105 \; & \; -4.526 \; & \; 3.367 \; & \; 2.614 \; & \; -2.164 \; & \; -3.056 \; & \; 1.977 \; & \; 0.800 \; & \; 2.187 \; \\
\; $F^{1^{+}}_3$ \; & \; 0.318 \; & \; 0.021 \; & \; 1.410 \; & \; 0.301 \; & \; 1.738 \; & \; 6.247 \; & \; 3.961 \; & \; 1.350 \; & \; 1.995 \; \\
\; $F^{0^{+} \leftrightarrow 1^{+}}$ \; & \; 0.869 \; & \; 0.113 \; & \; 1.201 \; & \; 0.033 \; & \; 1.783 \; & \; -5.225 \; & \; -2.330 \; & \; -1.757 \; & \; 1.848 \; \\
\hline\hline
\end{tabular}
\caption{\label{Vertex Dressing Funtion Parameters} Interpolation coefficients for each respective elastic photon+diquark form factor.}
\end{table*}

It is important to highlight that while the parameter $\eta$, which controls the strength of the AMM, must be completely determined by the quark-photon interaction itself~\cite{Xing:2021dwe}, the impact on the AMM could vary across the different diquark-photon scattering processes.


\subsubsection{The diagram $D_3$: photon-induced diquark transition}

The electro-coupling of a photon with a diquark of quantum numbers $J^P=0^+,1^+$ can induce a transition to a different diquark with $J'^{P'} =1^+,0^+$, given that in our approach we are only considering scalar ($sc$) and axial-vector ($ax$) diquarks. Electric charge conservation in $ax \leftrightarrow sc$ transitions entails a single electric charge $e_{J^P}$ characterizing this kind of processes. 

The corresponding scattering amplitude is represented in the lower diagram of Fig.~\ref{fig:EMInteractions} and defined mathematically by
\begin{eqnarray}
\Lambda^{D_3 \left( J^P \rightarrow J'^{P'} \right)}_{\mu \left[ \alpha \right] \left[ \beta \right]} \left(k; P_f, P_i\right) &=& \nonumber \\
&& \hspace{-3.7cm} e_{J^P} S(k) \mathcal{F}^{B_f}_{\left[ \alpha \right] \left[ \delta \right]} \Delta^{J'^{P'}}_{ \left[ \delta \rho \right]}(k_f) \Lambda_{\mu \left[ \rho \right] \left[ \sigma \right]}^{J^P \rightarrow J'^{P'}} \Delta^{J^P}_{ \left[ \sigma \lambda \right]}(k_i) \mathcal{F}^{B_i}_{\left[ \lambda \right] \left[ \beta \right]} \,, \nonumber \\
\end{eqnarray}
where the diquark-transition vertex, conveniently denoted as $\Lambda_{\mu \left[ \rho \right] \left[ \sigma \right]}^{J^P \rightarrow J'^{P'}} \equiv \Lambda_{\mu \left[ \rho \right] \left[ \sigma \right]}^{J^P \rightarrow J'^{P'}} \hspace{-.2cm} \left( Q, k^{1^{+}} \right)$, is given in terms of the axial-vector diquark momentum $k^{1^{+}}(=k_f, k_i)$ involved in the process, and defined as
\begin{eqnarray}
\Lambda_{\mu \left[ \rho \right] \left[ \sigma \right]}^{J^P \rightarrow J'^{P'}} \hspace{-.2cm} \left( Q, k^{1^{+}} \right) = F^{J^P \rightarrow J'^{P'}} \hspace{-.2cm} \left( Q^2 \right) \epsilon_{\mu \rho \alpha \beta} Q_{\alpha} k^{1^{+}}_{\beta} \,,
\end{eqnarray}
with the dressing function $F^{J^P \rightarrow J'^{P'}}$ separated again in terms of RP and AMM contributions, in analogy to the $F^{J^P}_{(j)}$ dressing functions for the elastic diquark-photon vertex counterparts (see Eq.~\eqref{Vertex Dressings decomposition}). The corresponding fit parameters are provided in the last row of Table~\ref{Vertex Dressing Funtion Parameters}, with identical values for $F^{0^{+} \rightarrow 1^{+}}$ and $F^{1^{+} \rightarrow 0^{+}}$.


\subsection{Elastic and transition electromagnetic currents}
\label{subsec:Currents}

The electromagnetic (EM) current of a nucleon can be expressed as
\begin{eqnarray}
\mathcal{J}_{\mu} \left( P_f, P_i \right) &=& \nonumber \\
&& \hspace{-1.8cm} i \Lambda_{+} \hspace{-.1cm} \left( P_f \right) \left[ F_1 \hspace{-.1cm} \left( Q^2 \right) \gamma_{\mu} +\frac{1}{2 m_N} F_2 \hspace{-.1cm} \left( Q^2 \right) \sigma_{\mu \nu} Q_{\nu} \right] \Lambda_{+} \hspace{-.1cm} \left( P_i \right) \,, \nonumber \\
\end{eqnarray}
where the dressing functions $F_1$ and $F_2$ are the Dirac and Pauli form factors, respectively. In order to connect with experiments, it is sometimes convenient to define the so-called electric and magnetic Sach form factors, respectively defined as
\begin{eqnarray}
G_E \left( Q^2 \right) &=& F_1 \hspace{-.1cm} \left( Q^2 \right) - \tau_N F_2 \hspace{-.1cm} \left( Q^2 \right)  \,, \label{GE Nucleon} \\
G_M \left( Q^2 \right) &=& F_1 \hspace{-.1cm} \left( Q^2 \right) + F_2 \hspace{-.1cm} \left( Q^2 \right)  \,,  \label{GM Nucleon}
\end{eqnarray}
with $\tau_N = Q^2/ \left[ 4m_N^2 \right]$. Moreover, the EM current must be normalized to ensure current conservation, this implies either $F_1(Q^2=0)=1$ or $G_E(Q^2=0)=1$, which in turn yields the normalization constant for the nucleon's Faddeev wave function. 

For Delta baryons, the general decomposition of the EM current can be expressed in terms of 4 form factors:
\begin{eqnarray}
\mathcal{J}_{\mu, \rho \sigma} \left( P_f, P_i \right) =  i \mathcal{P}^{\Delta}_{\rho \alpha}  \hspace{-.1cm} \left( P_f \right) \Gamma_{\mu, \alpha \beta} \left( P_f, P_i \right) \mathcal{P}^{N}_{\beta \sigma} \hspace{-.1cm} \left( P_i \right) \,,
\end{eqnarray}
where
\begin{eqnarray}
\Gamma_{\mu, \alpha \beta} \left( P_f, P_i \right) &=& \nonumber \\
&& \hspace{-2cm} \left[ F^{*}_1 \hspace{-.1cm} \left( Q^2 \right) \gamma_{\mu} +\frac{1}{2 m_N} F^{*}_2 \hspace{-.1cm} \left( Q^2 \right) \sigma_{\mu \nu} Q_{\nu} \right] \delta_{\alpha \beta} \nonumber \\
&& \hspace{-2cm} + 2\left[ F^{*}_3 \hspace{-.1cm} \left( Q^2 \right) \gamma_{\mu} +\frac{1}{2 m_N} F_4^{*} \hspace{-.1cm} \left( Q^2 \right) \sigma_{\mu \nu} Q_{\nu} \right] \frac{Q_{\alpha} Q_{\beta}}{4 m_{\Delta}^2} \,. \nonumber \\
\label{Delta-Photon Vertex} 
\end{eqnarray}

In connection with the ground-state Delta baryon, experimental data have usually provided insights into the contributions of various photon multipoles, namely the electric charge, $G_{E0}$, the magnetic dipole, $G_{M1}$, the electric quadrupole, $G_{E2}$, and the magnetic octupole, $G_{M3}$. These contributions are linearly related to the conventional $F_i^\ast$, with $i=1,\ldots,4$, form factors as~\cite{Segovia:2013uga} 
\begin{eqnarray}
G_{E0} \hspace{-.1cm} \left( Q^2 \right) \hspace{-.2cm} &=& \hspace{-.2cm} \left( 1+ \frac{2 \tau_{\Delta}}{3} \right) \Delta F^{*}_{1,2}   - \frac{\tau_{\Delta}}{3} \left( 1+ \tau_{\Delta} \right) \Delta F^{*}_{3,4} \,, \hspace{.9cm}  \label{GE0 Delta} \\
G_{M1} \hspace{-.1cm} \left( Q^2 \right) \hspace{-.2cm} &=& \hspace{-.2cm} \left( 1+ \frac{4 \tau_{\Delta}}{5} \right) \Sigma F^{*}_{1,2}  - \frac{2 \tau_{\Delta}}{5} \left( 1+ \tau_{\Delta} \right) \Sigma F^{*}_{3,4} \,, \label{GM1 Delta} \\
G_{E2} \hspace{-.1cm} \left( Q^2 \right) \hspace{-.2cm} &=& \hspace{-.2cm} \Delta F^{*}_{1,2}  - \frac{1}{2} \left( 1+ \tau_{\Delta} \right) \Delta F^{*}_{3,4} \,, \label{GE2 Delta} \\
G_{M3} \hspace{-.1cm} \left( Q^2 \right) \hspace{-.2cm} &=& \hspace{-.2cm} \Sigma F^{*}_{1,2} - \frac{1}{2} \left( 1+ \tau_{\Delta} \right) \Sigma F^{*}_{3,4} \,, \label{GM3 Delta}
\end{eqnarray}
with $\Sigma F^{*}_{i,j} = F^{*}_i + F^{*}_j $, $\Delta F^{*}_{i,j} = F^{*}_i - \tau_{\Delta} F^{*}_j $ and $\tau_{\Delta} = Q^2/ \left[ 4m_{\Delta}^2 \right]$. It is important to stress again that current conservation demands that either $F_1^\ast$ or $G_{E0}$ are equal to $1$ at $Q^2=0\,\text{GeV}^2$, which in turn entail the normalization constant for the $\Delta$-baryon.

The photon-induced transition current of a Nucleon to a Delta-baryon can be represented quite generally as 
\begin{eqnarray}
\mathcal{J}_{\mu, \rho} \left( P_f, P_i \right) =  i \mathcal{P}^{\Delta}_{\rho \alpha}  \hspace{-.1cm} \left( P_f \right) \Gamma_{\mu, \alpha} \left( P_f, P_i \right) \Lambda_{+} \hspace{-.1cm} \left( P_i \right) \,, \label{Transition current}
\end{eqnarray}
where the decomposition for $\Gamma_{\mu, \alpha} \left( P_f, P_i \right)$ can be written in terms of three dressing functions: the so-called Jones-Scadron form factors corresponding to the magnetic dipole, $G^{*}_M$, the electric quadrupole, $G^{*}_E$, and the Coulomb quadrupole, $G^{*}_C$:
\begin{eqnarray}
\Gamma_{\mu, \alpha} \left( P_f, P_i \right) &=& \nonumber \\
&& \hspace{-2cm} b \left[ -\frac{i \omega}{2 \lambda_{+}} \left( G^{*}_M - G^{*}_E \right) \mathcal{V}^1_{\alpha \mu} - G^{*}_E \mathcal{V}^2_{\alpha \mu} + \frac{i \tau}{\omega} G^{*}_C \mathcal{V}^3_{\alpha \mu} \right] \,, \nonumber \\ \label{Transition vertex}
\end{eqnarray}
where
\begin{eqnarray}
\lambda_{+} &=& \frac{1+\sqrt{1-4\delta^2}}{2} + \tau \,, \\
\omega &=& \sqrt{\delta^2 + \tau \left( 1 + \tau \right)} \,, \\
b &=& \sqrt{\frac{3}{2}} \left( 1+ \frac{m_{\Delta}}{m_N} \right) \,.
\end{eqnarray}
with $\tau = Q^2 \hspace{-.05cm}/ \hspace{-.1cm} \left[ 4 \bar{m}^2 \right]$, $\delta = \left( m_{\Delta}^2-m_N^2\right)\hspace{-.1cm} / \hspace{-.1cm}\left[ 4 \bar{m}^2 \right]$ and  $\bar{m}^2 = \left( m_N^2 + m_{\Delta}^2 \right)\hspace{-.1cm}/2$. Moreover, the tensors $\mathcal{V}^i_{\alpha \mu} \equiv \mathcal{V}^i_{\alpha \mu} \left( K, Q \right)$ are defined as ($2 K \equiv P_f + P_i$)
\begin{eqnarray}
\mathcal{V}^1_{\alpha \mu} \left( K, Q \right) &=&  \gamma_5 \epsilon_{\alpha \mu \sigma \rho} \mathcal{K}_{\sigma} \left( K,Q \right) \hat{Q}_{\rho} \,, \\
\mathcal{V}^2_{\alpha \mu} \left( K, Q \right) &=& \left( \delta_{\alpha \sigma} - \hat{Q}_{\alpha} \hat{Q}_{\sigma}\right) \bar{\mathcal{T}}_{\sigma \mu} \left( K, Q\right) \,, \\
\mathcal{V}^3_{\alpha \mu} \left( K, Q \right) &=&  \hat{Q}_{\alpha} \mathcal{K}_{\mu} \left( K,Q \right) \,,
\end{eqnarray}
with $\hat{Q} = Q/\left[ 2 \bar{m} \tau \right]$ and
\begin{eqnarray}
\mathcal{K}_{\sigma} \left( K, Q \right) &\equiv& \frac{\sqrt{\tau}}{i \bar{m} \omega} \left( K_{\sigma} + \frac{\delta}{2 \tau} Q_{\sigma} \right) \,, \\
\bar{\mathcal{T}}_{\sigma \mu} \left( K, Q \right) &\equiv& \delta_{\sigma \mu} - \mathcal{K}_{\sigma} \left( K, Q \right)  \mathcal{K}_{\mu} \left( K, Q \right) \,.
\end{eqnarray}

In order to compare directly with experimental measurements, it is often necessary to compute the so-called helicity amplitudes for the transition. For the $\gamma^{(*)} + N(940)\frac{1}{2}^+ \rightarrow \Delta(1700)\frac{3}{2}^{-}$ reaction, they are expressed in terms of the Jones-Scadron form factors as follows
\begin{eqnarray}
\mathcal{A}_{1/2} \left( Q^2 \right) &=& -\frac{1}{4 \mathcal{F}} \left[ G^{*}_{E} \left( Q^2 \right) - 3 G^{*}_{M} \left( Q^2 \right) \right] \,, \label{Helicity Amplitude A12} \\
\mathcal{A}_{3/2} \left( Q^2 \right) &=& -\frac{\sqrt{3}}{4 \mathcal{F}} \left[ G^{*}_{E} \left( Q^2 \right) + G^{*}_{M} \left( Q^2 \right) \right] \,, \label{Helicity Amplitude A32} \\
\mathcal{S}_{1/2} \left( Q^2 \right) &=& -\frac{1}{\sqrt{2} \mathcal{F}} \frac{|\mathbf{q}|}{2 m_{\Delta}} G^{*}_{C} \left( Q^2 \right)\,, \label{Helicity Amplitude S12} 
\end{eqnarray}
where $|\bm{q}|=\sqrt{Q_+^2Q_-^2}/(2m_\Delta)$, with $Q_{\pm}^2=(m_{\Delta}\pm m_N)^2+Q^2$, and
\begin{eqnarray}
\mathcal{F} = \frac{1}{\sqrt{2 \pi \alpha}} \frac{m_N}{m_{\Delta} - m_N} \sqrt{\frac{m_N \left( m^2_{\Delta} - m^2_N \right)}{\left( m_{\Delta} + m_N\right)^2 +Q^2 }} \,,
\end{eqnarray}
with $\alpha = 1/137$. In the next section, we shall provide results for the computed Jones-Scadron form factors as well as their associated helicity amplitudes, for the $\gamma^{(*)} + N(940)\frac{1}{2}^+ \rightarrow \Delta(1700)\frac{3}{2}^{-}$ transition.  


\section{RESULTS}
\label{sec:Results}

A detailed discussion of the intricacies of our numerical results for the transition form factors, and then for the corresponding helicity amplitudes, that describe the $\gamma^{(*)} + N(940)\frac{1}{2}^+ \to \Delta(1700)\frac{3}{2}^{-}$ reaction shall be given below. However, a few comments are in order first.

The first one refers to the SCI-DSEs framework and its parameters. Note that the parameters of the model used herein have already been constrained in earlier works. Consequently, from this perspective, we introduce no further parameters to study the transition form factors presented in this work. The second comment is then related to our theoretical uncertainties. There are two types of them: one is intrinsic to the numerical algorithm and the other is related to the way the parameters are fixed. The numerical error is negligible and, as mentioned above, the model parameters are adjusted to reproduce a certain small number of hadron observables within an acceptable range of agreement with experiment. Therefore, it is not illustrative to assign an error to these parameters and consequently to the observable quantities inferred from them. 

In order to analyze the associated parameter uncertainty of the calculations presented in this manuscript, the numerical findings are represented as a colored band, which reflects the variation of the parameter $\eta$ governing the dressed-quark anomalous magnetic moment. Specifically, the solid line corresponds to $\eta = 0$, the dotted line to $\eta = 2/3$, and the dashed line serves as a reference, representing the central value computed for $\eta = 1/3$. This parametrization provides a comprehensive analysis of the sensitivity of our results to the quark dynamics in the transition form factors. Another analysis of theoretical uncertainties that we shall perform is related with the dependence of our computed results with respect to the masses of $N(940)\frac{1}{2}^+$ and $\Delta(1700)\frac{3}{2}^-$ baryons. This variation effectively mimics the beyond-rainbow-ladder effects, such as the pion-cloud which is known to contribute up to about 20\% to the form factors for low momentum transfer.

\begin{figure}[!t]
\centering
\includegraphics[scale=.25]{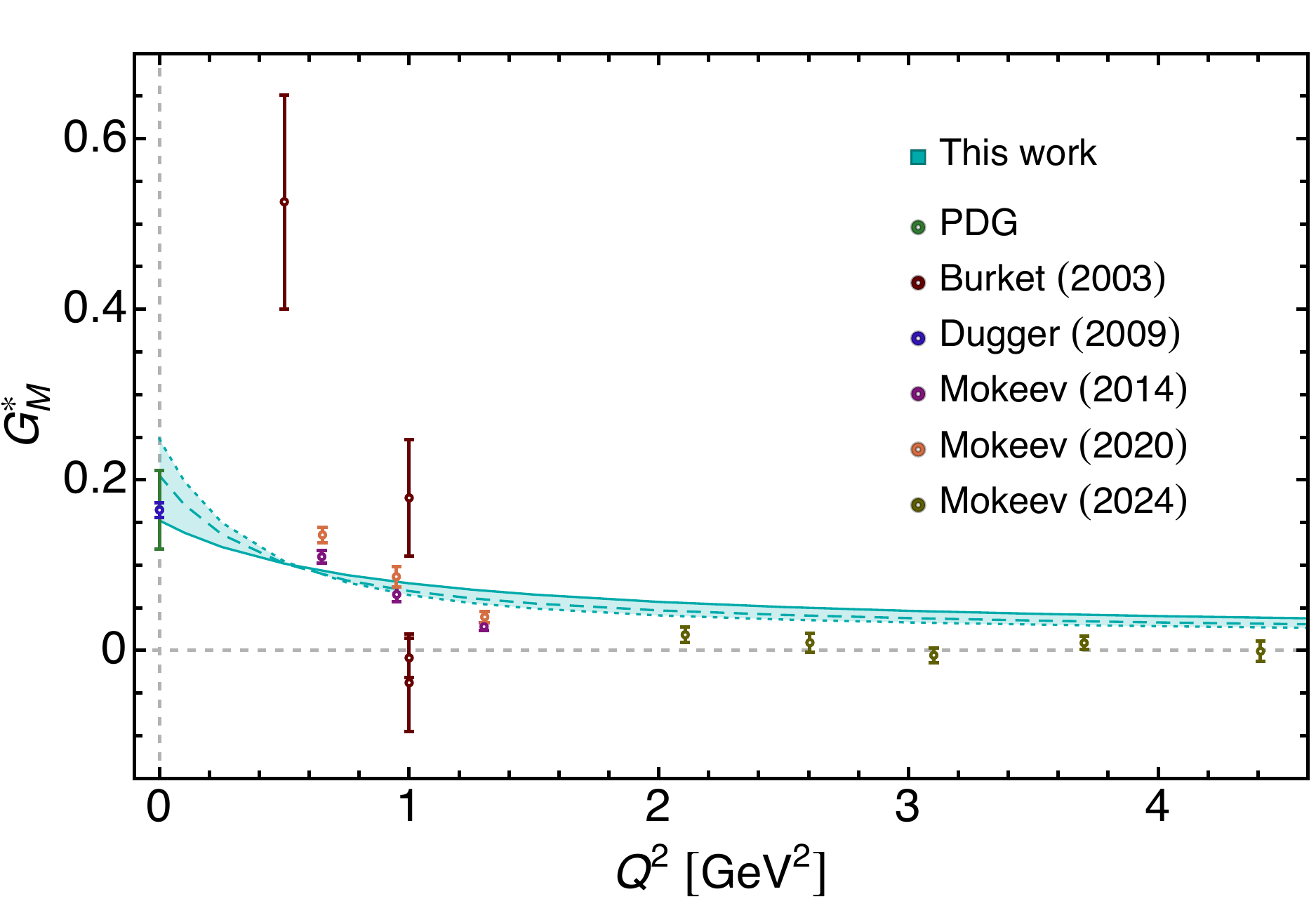}
\caption{\label{GM} Magnetic dipole ($G^{*}_M$) transition form factor as a function of $Q^2$. The data points correspond to experimental measurements taken from Refs.~\cite{ParticleDataGroup:2024cfk, Burkert:2002zz, CLAS:2009tyz, Mokeev:2013kka, Mokeev:2020hhu, Mokeev:2024beb}. The cyan band represents our numerical results obtained for different values of the parameter $\eta$ that modulates the AMM: $\eta = 0$ (solid line), $\eta = 1/3$ (dashed line), and $\eta = 2/3$ (dotted line).}
\end{figure}

\begin{figure}[!t]
\centering
\includegraphics[scale=.25]{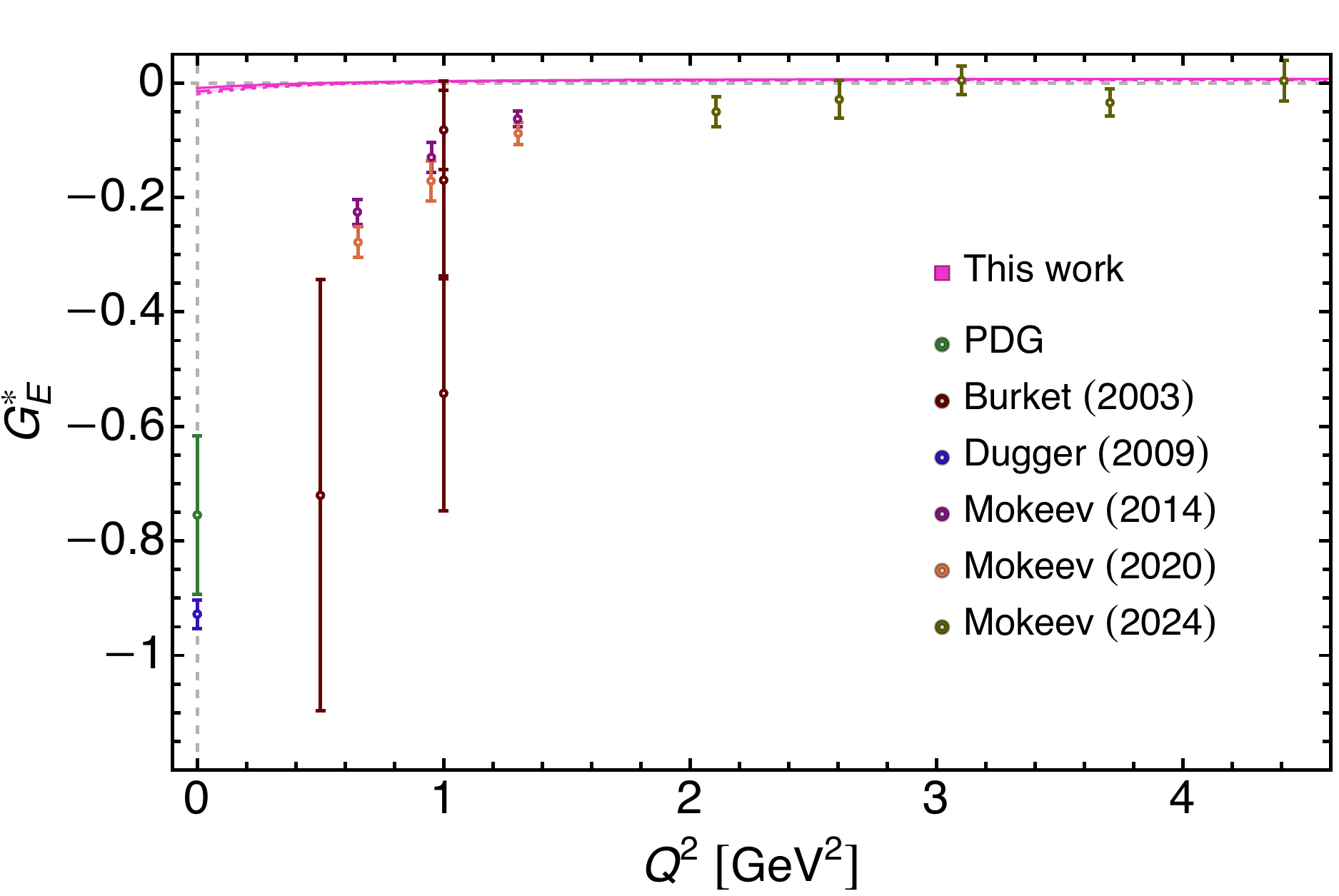}
\caption{\label{GE} Electric quadrupole ($G^{*}_E$) transition form factor as a function of $Q^2$. The data points correspond to experimental measurements taken from Refs.~\cite{ParticleDataGroup:2024cfk, Burkert:2002zz, CLAS:2009tyz, Mokeev:2013kka, Mokeev:2020hhu, Mokeev:2024beb}. The magenta band represents our numerical results obtained for different values of the parameter $\eta$ that modulates the AMM: $\eta = 0$ (solid line), $\eta = 1/3$ (dashed line), and $\eta = 2/3$ (dotted line).}
\end{figure}

\begin{figure}[!t]
\centering
\includegraphics[scale=.25]{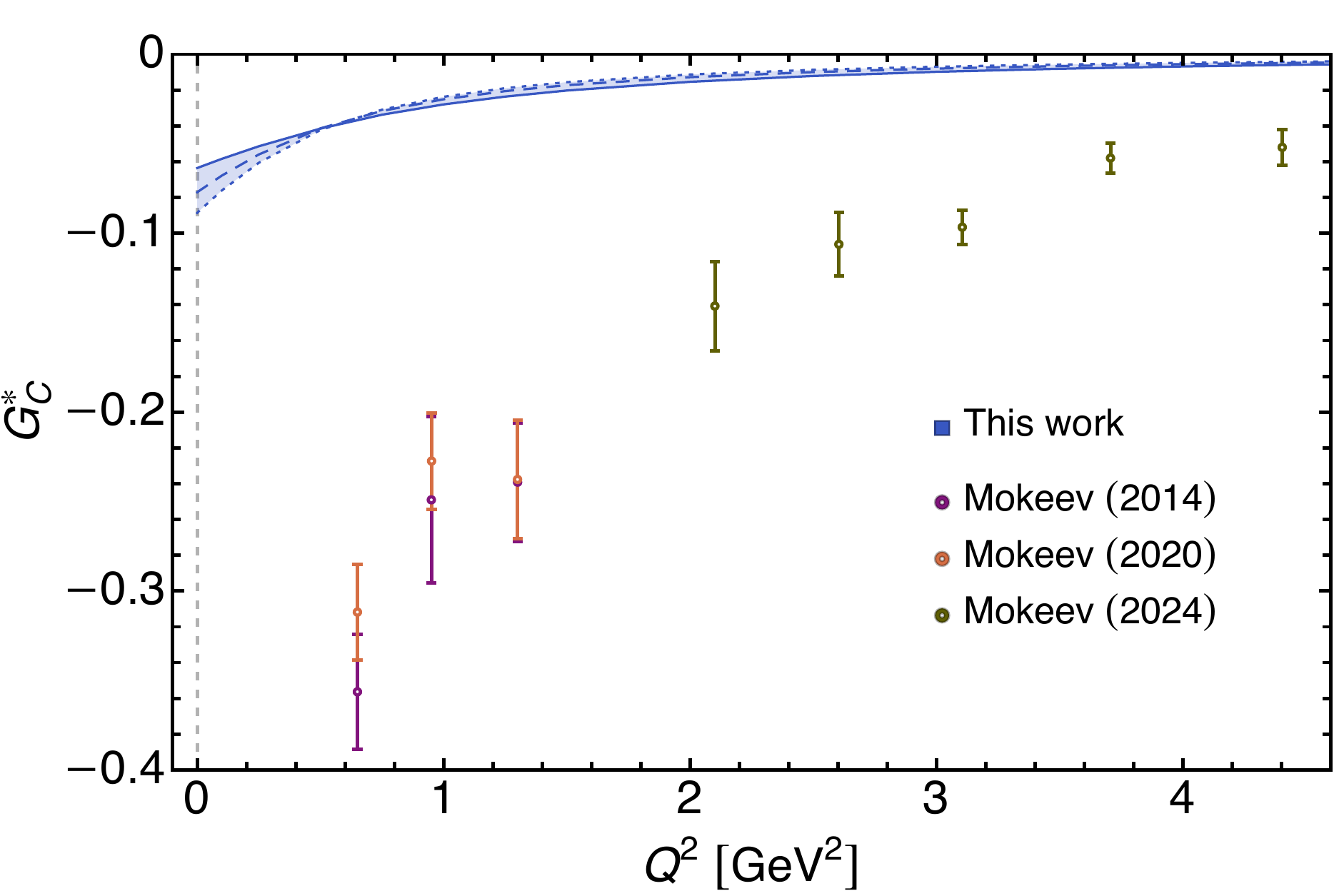}
\caption{\label{GC} Coulomb quadrupole ($G^{*}_C$) transition form factor as a function of $Q^2$. The data points correspond to experimental measurements taken from Refs.~\cite{ParticleDataGroup:2024cfk, Burkert:2002zz, CLAS:2009tyz, Mokeev:2013kka, Mokeev:2020hhu, Mokeev:2024beb}. The blue band represents our numerical results obtained for different values of the parameter $\eta$ that modulates the AMM: $\eta = 0$ (solid line), $\eta = 1/3$ (dashed line), and $\eta = 2/3$ (dotted line).}   
\end{figure}

The numerical results for the magnetic dipole ($G^{*}_M$), electric quadrupole ($G^{*}_E$), and Coulomb ($G^{*}_C$) quadrupole transition form factors, which define the transition current $\gamma^{(*)} + N(940)\frac{1}{2}^+ \rightarrow \Delta(1700)\frac{3}{2}^{-}$, are shown in Figs.~\ref{GM} to \ref{GC}, and are also compared with the available experimental data~\cite{ParticleDataGroup:2024cfk, Burkert:2002zz, CLAS:2009tyz, Mokeev:2013kka, Mokeev:2020hhu, Mokeev:2024beb}.

The magnetic dipole transition form factor, $G^{*}_M$, is shown in Fig.~\ref{GM} as a function of $Q^2$. The cyan band represents our numerical results within the SCI approach, with its width reflecting variations in the model parameter $\eta \in \left[ 0, 2/3 \right]$. At $Q^2=0$, our results are in excellent agreement with the experimental values, lying well within the experimental error bands. In the intermediate region, $0.6 \, \hbox{GeV}^2 < Q^2 < 1.4 \, \hbox{GeV}^2$, the computed form factor closely matches experimental data, underscoring the robustness of our approach. At higher momentum transfers $Q^2 > 1.4 \, \text{GeV}^2$, our computation produces a gradual, smooth decline, while experimental data indicate a sharper drop-off in this region. Nonetheless, the overall agreement emphasizes the reliability of the SCI framework for describing the leading magnetic form factor of the transition.

In Fig.~\ref{GE}, we present the results for the electric quadrupole transition form factor $G^{*}_E$ as a function of $Q^2$. The thin magenta band represents our numerical results for $\eta \in \left[ 0, 2/3 \right]$. At large $Q^2$, our theoretical curve exhibits a sign flip, transitioning from negative to positive values, consistent with the trend observed in the experimental data. This sign change may be a notable feature of the dynamical behavior in the $\gamma^{(*)} + N(940)\frac{1}{2}^+ \to \Delta(1700)\frac{3}{2}^{-}$ transition process. In the low $Q^2$-regime ($Q^2 < 1\,\text{GeV}^2$), while both our results and the experimental data remain negative, a significant discrepancy arises: the computed values stay relatively small and stable, whereas the experimental data show a dramatic fall as $Q^2 \rightarrow 0$. This sharp decline in experimental values highlights the enhanced role of contributions to the electric quadrupole transition form factor at low momenta, which are not fully captured within the SCI approach. Specifically, since our theoretical result for $G_M^\ast$ is reasonably good at low values of transferred momenta, the underestimation of quark orbital angular momentum content within the bound-state baryons because having Faddeev amplitudes that are independent of relative momentum could explain small values for higher order electromagnetic multipoles, such as the $G_E^\ast$ and $G_C^\ast$ in this transition.

Finally, our results for the Coulomb quadrupole transition form factor $G^{*}_C$ are presented in Fig.~\ref{GC}, and compared with the available experimental data. The blue band displays our numerical outcome for $\eta \in \left[ 0, 2/3 \right]$. Notably, there is no experimental data available for $Q^2 < 0.6~\text{GeV}^2$. While our calculations and the experimental data agree on the sign of the form factor, a significant discrepancy is observed: our computed values remain consistently smaller than the experimental results. Again, this difference suggests that important dynamical contributions are missing in the current theoretical approach. In particular, a SCI which produces Faddeev amplitudes that are independent of relative momentum must underestimate the quark orbital angular momentum content of the bound-state, which are crucial to explain higher order multipole contributions of electromagnetic transitions.

\begin{figure}[!t]
\centering
\includegraphics[scale=.25]{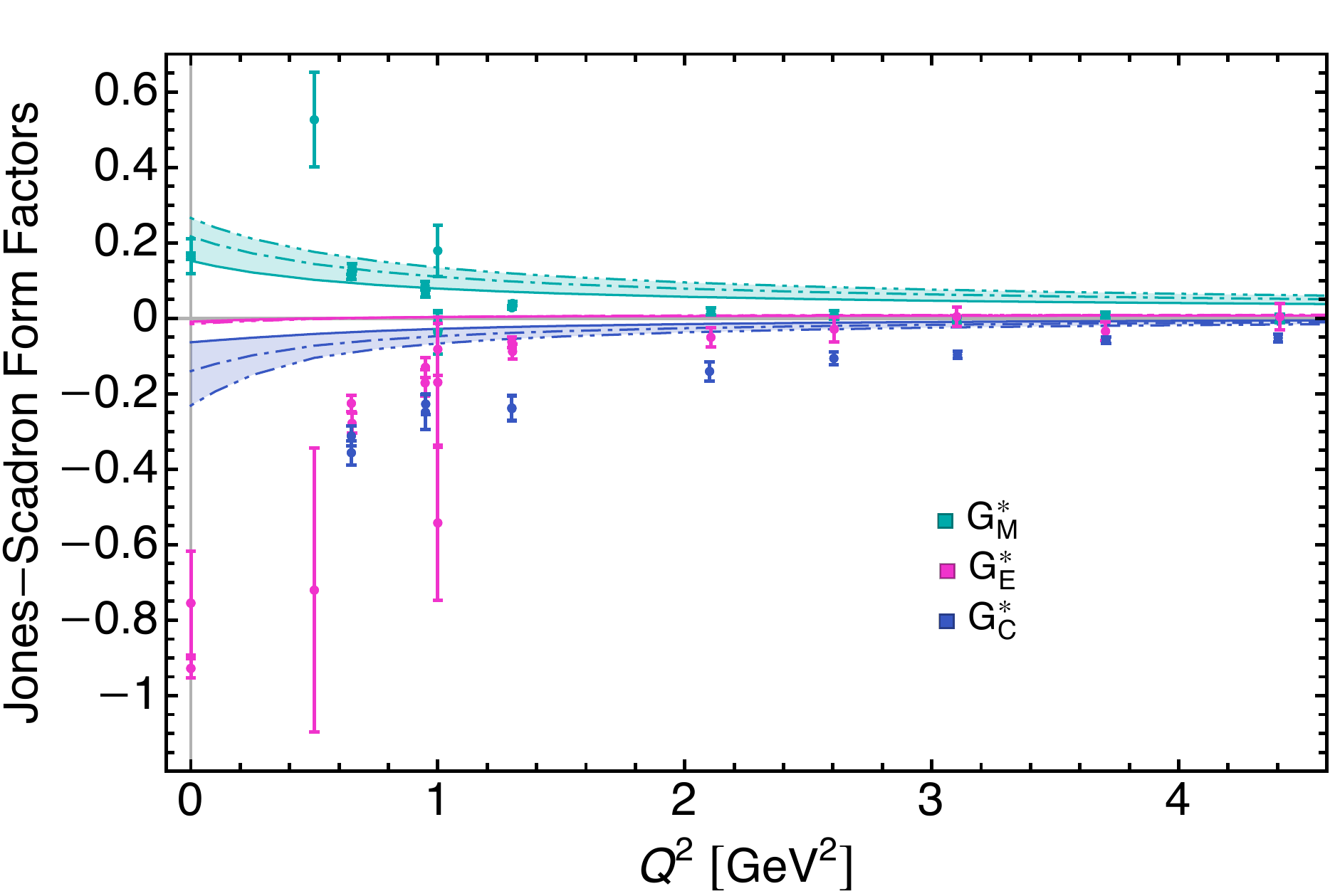}
\caption{\label{Jones-Scadron FFs - Delta mass variation} Qualitative analysis of the sensitivity of electromagnetic transition form factors $G_M^\ast$, $G_E^\ast$ and $G_C^\ast$ of the $\gamma^{(\ast)} + N(940)\frac{1}{2}^+\to \Delta(1700)\frac{3}{2}^-$ reaction with respect to the mass of the $\Delta(1700)\frac{3}{2}^-$ baryon. For each colored band, the solid line represents $m_{\Delta} = 1.72 \, \text{GeV}$ calculation, the dashed-dotted line corresponds to $m_{\Delta} = 1.80 \, \text{GeV}$, and the dashed-double-dotted line stands for $m_{\Delta} = 1.88 \, \text{GeV}$.}
\end{figure}

In order to roughly illustrate the sensitivity of the transition form factors with respect to beyond-rainbow-ladder truncation, we show in Fig.~\ref{Jones-Scadron FFs - Delta mass variation}, for $\eta = 0$, their variation with different values of the Delta's mass. Each solid line represents $m_{\Delta} = 1.72 \, \text{GeV}$ calculation, the dash-dot line corresponds to $m_{\Delta} = 1.80 \, \text{GeV}$, and the dash-double-dot line stands for $m_{\Delta} = 1.88 \, \text{GeV}$. Our results show that the form factors become infrared enhanced as the $\Delta$ baryon mass increases, which goes in the right direction in order to get a better agreement with experimental data for $G_E^\ast$ and $G_C^\ast$ at low $Q^2$. Nonetheless, such sensitivity is very moderate and the results exhibit a convergence to the original computation at high values of transferred momenta.

\begin{figure}[!t]
\centering
\includegraphics[scale=.25]{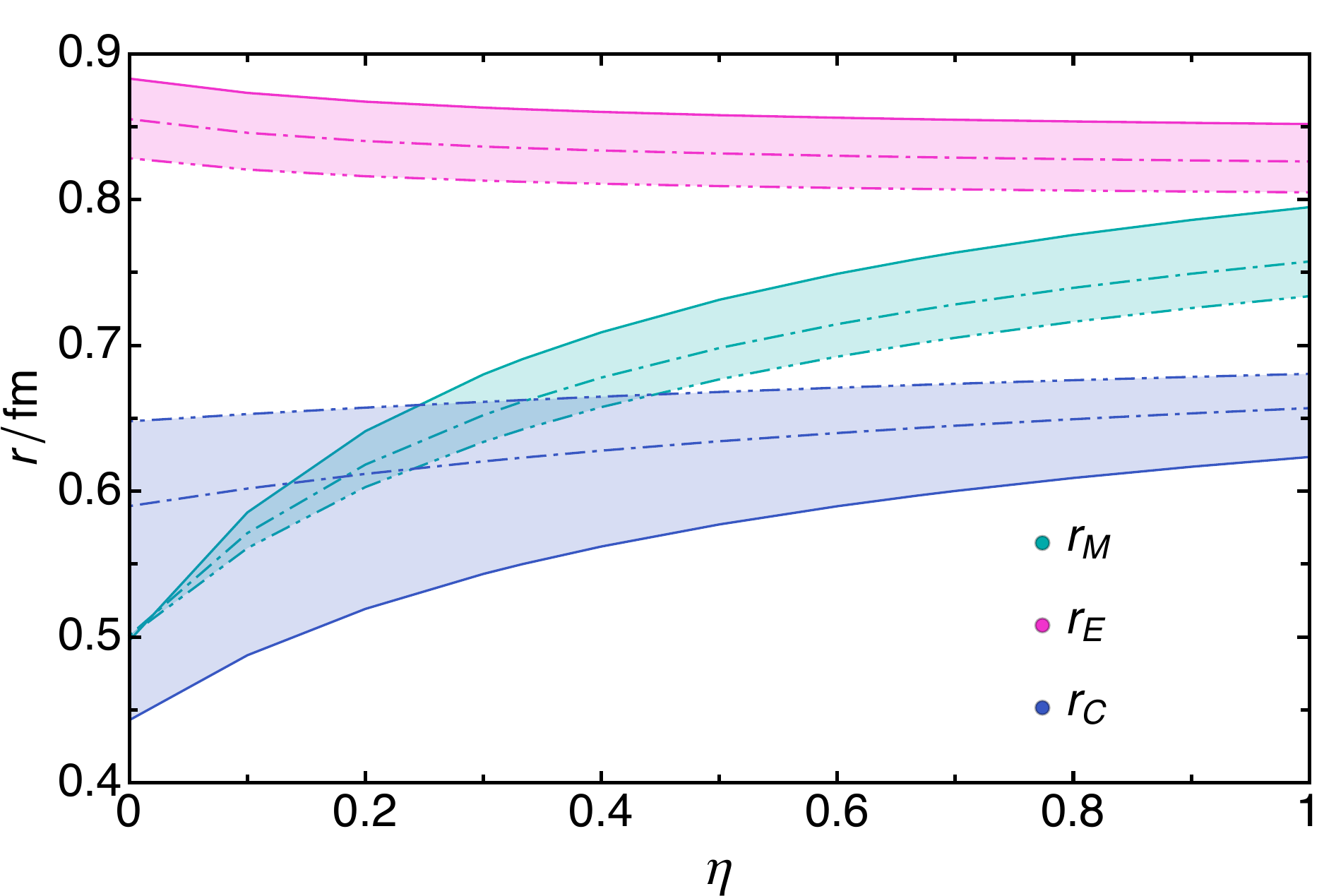}
\caption{\label{Radii} The magnetic dipole ($r_M$), electric quadrupole ($r_E$), and Coulomb quadrupole ($r_C$) square-root of mean square radii as a function of the anomalous magnetic moment of the dressed quark. For each colored band, the solid line represents $m_{\Delta} = 1.72 \, \text{GeV}$ calculation, the dashed-dotted line corresponds to $m_{\Delta} = 1.80 \, \text{GeV}$, and the dashed-double-dotted line stands for $m_{\Delta} = 1.88 \, \text{GeV}$.}
\end{figure}

For completeness, we compute the magnetic dipole ($r_M$), electric quadrupole ($r_E$), and Coulomb quadrupole ($r_C$) square root of mean square radii, defined as 
\begin{eqnarray}
r^2_{\mathcal{F}} =  - \frac{6}{\mathcal{F}(0)} \left. \frac{d \mathcal{F}(Q^2)}{d Q^2} \right|_{Q^2 = 0} \,,
\end{eqnarray}
for the associated form factors $\mathcal{F} = G^{*}_{M}, \, G^{*}_{E}, \, G^{*}_{C}$, respectivly. Figure~\ref{Radii} illustrates the sensitivity of the computed radii to the anomalous magnetic moment of the dressed quark. Our findings indicate that each radius converges to a stable value as $\eta$ increases. Notably, Table~\ref{Radii Table} provides the explicit values of the radii calculated at the central parameter value, $\eta = 1/3$, with the uncertainties defined as the difference between the corresponding radius' values at $\eta = 0$ and $\eta = 2/3$. Note also in the Table~\ref{Radii Table} that the radii exhibit a pattern of increasing values with a higher $\Delta(1700)\frac{3}{2}^-$ baryon mass.

\begin{table}[!t]
\centering
\renewcommand{\arraystretch}{1.5}
\begin{tabular}{|c|c|c|c|}
\hline
 \; $m_{\Delta}/$GeV \; & 1.72 & 1.80 & 1.88 \\
\hline
$r_{M}$/fm & \; $0.69_{-0.19}^{+0.07}$ \; & \; $0.66_{-0.16}^{+0.06}$ \; & \; $0.64_{-0.14}^{+0.06}$ \; \\
$r_{E}$/fm & $0.86_{-0.01}^{+0.02}$ & $0.84_{-0.01}^{+0.02}$ & $0.81_{-0.01}^{+0.02}$ \\
$r_{C}$/fm & $0.55_{-0.11}^{+0.05}$ & $0.62_{-0.03}^{+0.02}$ & $0.66_{-0.01}^{+0.01}$ \\
\hline
\end{tabular}
\caption{\label{Radii Table} The magnetic dipole ($r_M$), electric quadrupole ($r_E$), and Coulomb quadrupole ($r_C$) square-root of mean square radii as a function of the anomalous magnetic moment of the dressed quark (error bars) and the mass of the $\Delta(1700)\frac{3}{2}^-$. See related text for details.}
\end{table}

\begin{figure}[!t]
\centering
\includegraphics[scale=.25]{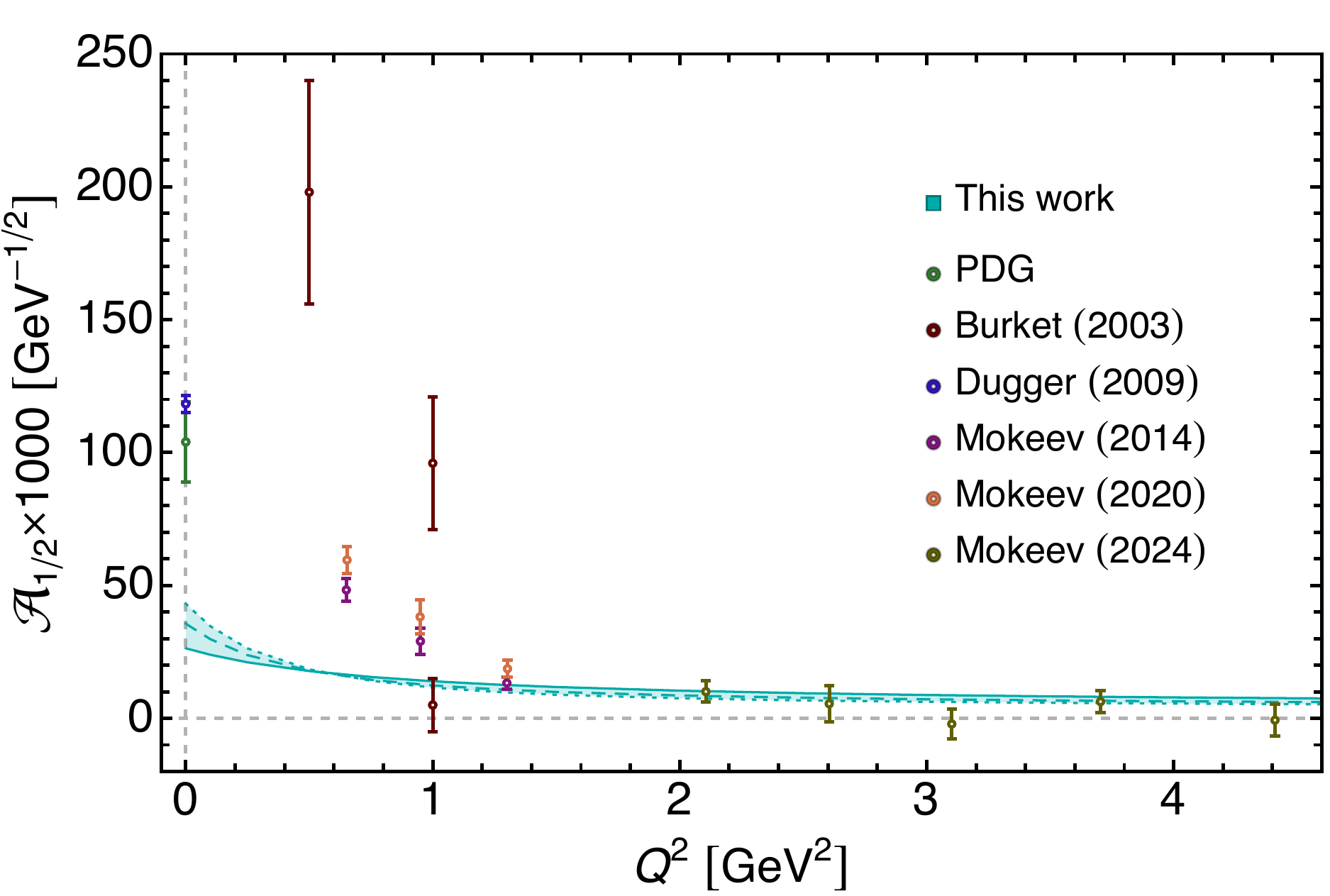}
\includegraphics[scale=.25]{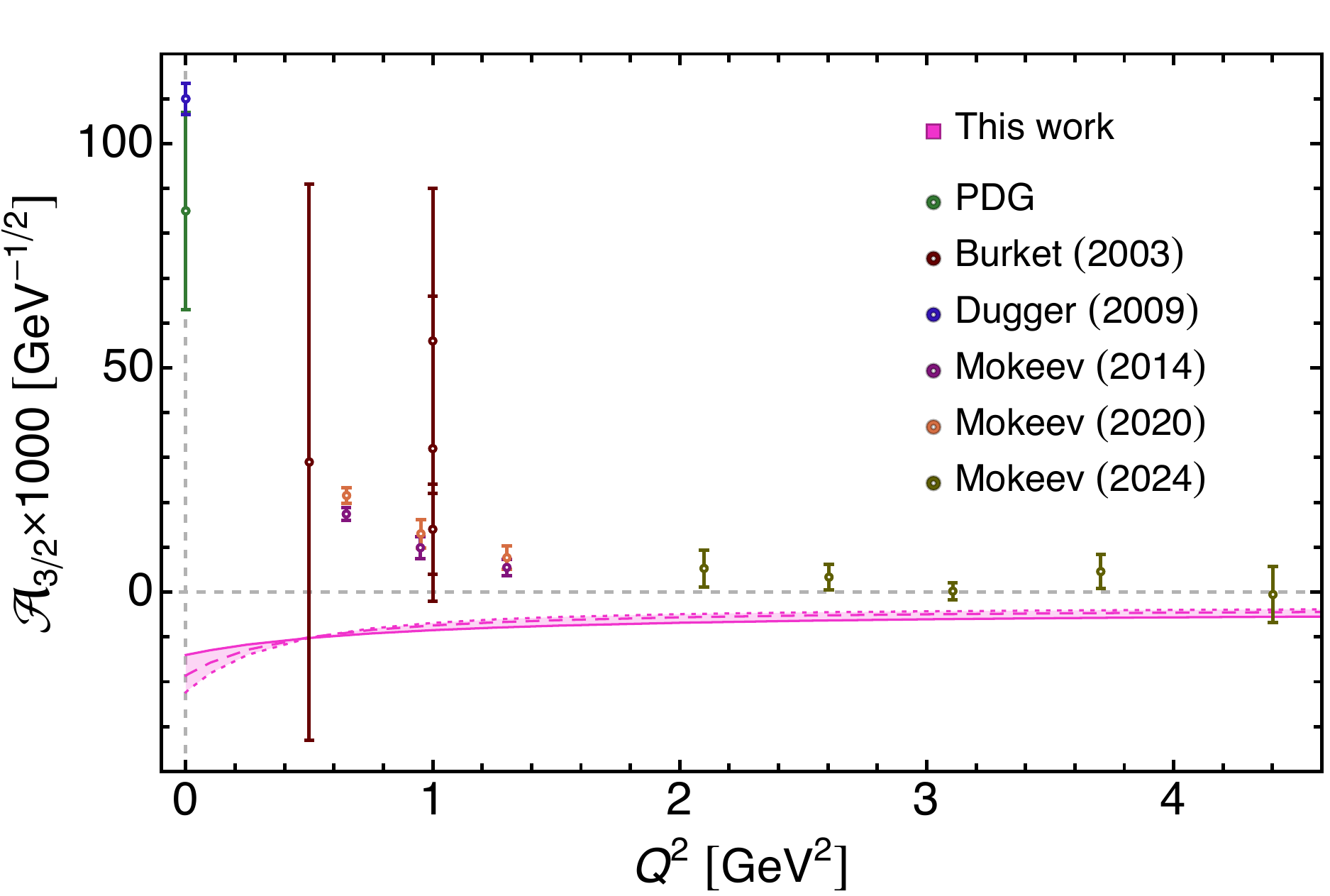}
\includegraphics[scale=.25]{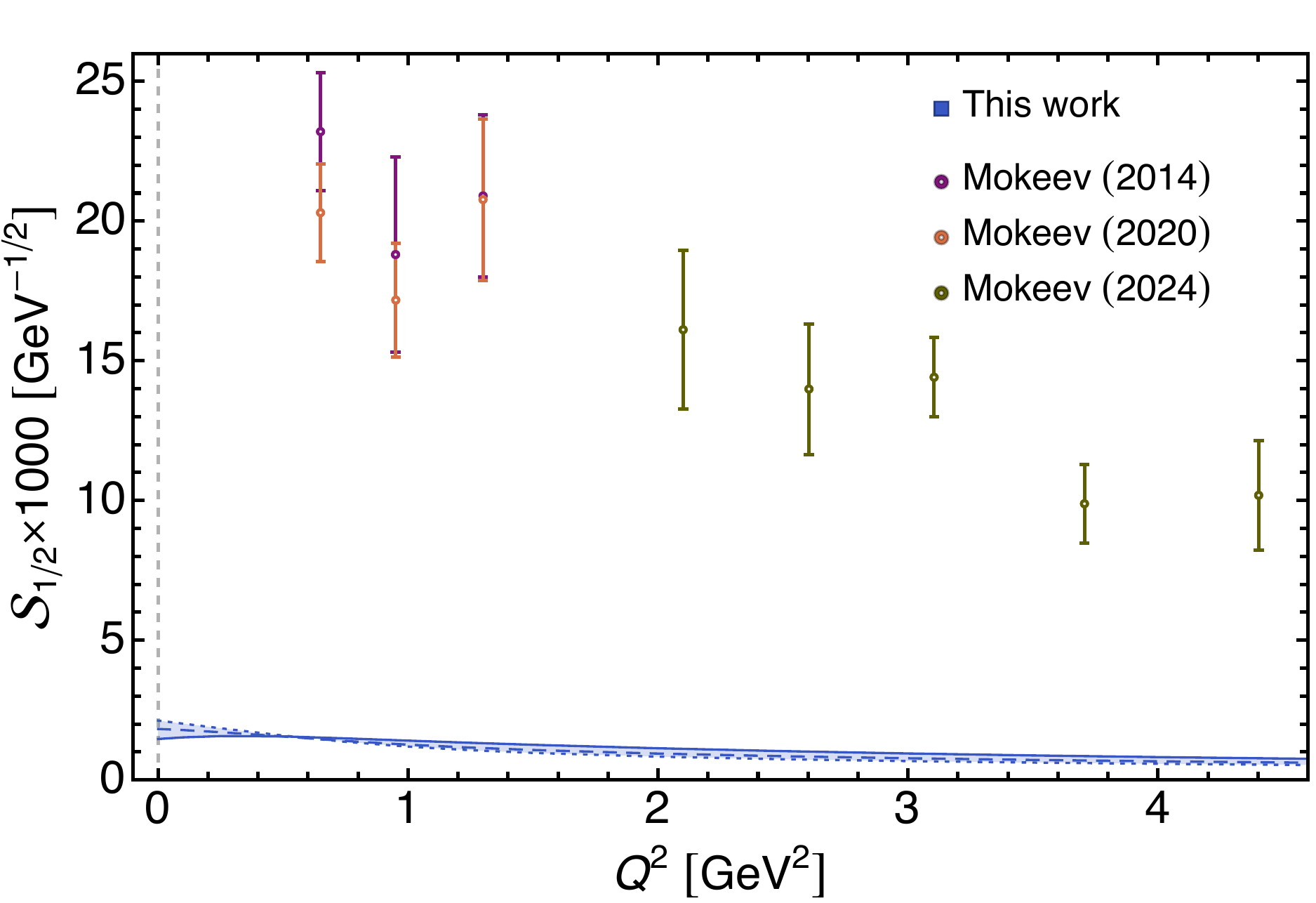}
\caption{\label{fig:helicities} Helicity amplitudes $\mathcal{A}_{1/2}$ (top panel), $\mathcal{A}_{3/2}$ (middle panel) and $\mathcal{S}_{1/2}$ (bottom panel) as a function of $Q^2$. In each panel, the band represents our numerical results obtained for different values of the parameter $\eta$ that modulates the AMM: $\eta = 0$ (solid line), $\eta = 1/3$ (dashed line), and $\eta = 2/3$ (dotted line). The experimental data is taken from Refs.~\cite{ParticleDataGroup:2024cfk, Burkert:2002zz, CLAS:2009tyz, Mokeev:2013kka, Mokeev:2020hhu, Mokeev:2024beb}.}
\end{figure}

Figure~\ref{fig:helicities} presents analogous results for the helicity amplitudes $\mathcal{A}_{1/2}$ (top panel), $\mathcal{A}_{3/2}$ (middle panel) and $\mathcal{S}_{1/2}$ (bottom panel), as a function of $Q^2$, that describes the $\gamma^{(\ast)} + N(940)\frac{1}{2}^+\to \Delta(1700)\frac{3}{2}^-$ transition. These results are obtained from the transition form factors $G^{*}_{M}$, $G^{*}_{E}$ and $G_C^\ast$ by using Eqs.~\eqref{Helicity Amplitude A12} to~\eqref{Helicity Amplitude S12}. From the top panel of Fig.~\ref{fig:helicities}, one can see that our computed values for $\mathcal{A}_{1/2}$ align well with experimental data for $Q^2 > 1~\text{GeV}^2$; however, they exhibit a weaker enhancement at lower $Q^2$ compared to the rapid rise observed in the experimental results. This mismatch is connected to small values of $G_E^\ast$ at low transferred momenta and highlights the need for additional contributions or refinements in the theoretical framework at low $Q^2$. The helicity amplitude $\mathcal{A}_{3/2}$, shown in the middle panel of Fig.~\ref{fig:helicities}, displays a notable feature: a sign flip is observed in our results compared to the experimental data. This disparity can be again attributed to the small values of the $G^{*}_{E}$ transition form factor obtained within the SCI approach. In the bottom panel of Fig.~\ref{fig:helicities}, we show our results for $\mathcal{S}_{1/2}$ which matches the sign of the experimental data but remains at lower values compared to the data for entire domain of $Q^2$ spanned. This behavior was anticipated due to the discrepancy observed in the Coulomb transition form factor.

\begin{figure}[!t]
\centering
\includegraphics[scale=.25]{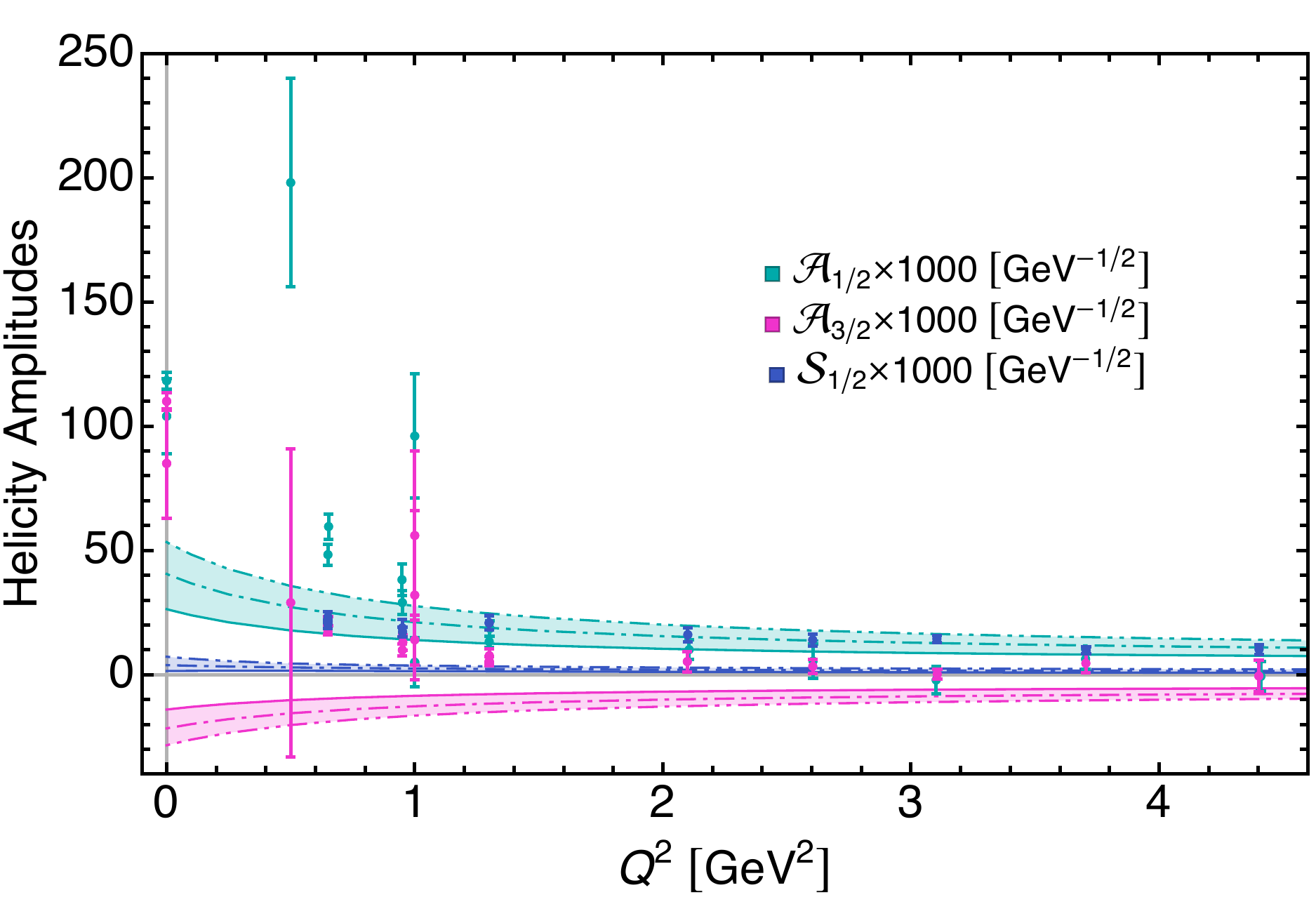}
\caption{\label{Helicity Amplitudes - Delta Mass Variation} Qualitative analysis of the sensitivity of helicity amplitudes $\mathcal{A}_{1/2}$, $\mathcal{A}_{3/2}$ and $\mathcal{S}_{1/2}$ of the $\gamma^{(\ast)} + N(940)\frac{1}{2}^+\to \Delta(1700)\frac{3}{2}^-$ reaction with respect to the mass of the $\Delta(1700)\frac{3}{2}^-$ baryon. For each colored band, the solid line represents $m_{\Delta} = 1.72 \, \text{GeV}$ calculation, the dashed-dotted line corresponds to $m_{\Delta} = 1.80 \, \text{GeV}$, and the dashed-double-dotted line stands for $m_{\Delta} = 1.88 \, \text{GeV}$.}
\end{figure}

To provide a qualitative understanding of how the helicity amplitudes respond to beyond-rainbow-ladder truncation effects, Fig.~\ref{Helicity Amplitudes - Delta Mass Variation} illustrates their dependence on varying values of the Delta baryon mass for $\eta = 0$. For each colored band, the solid line corresponds to calculations with $m_\Delta = 1.72 \, \text{GeV}$, the dash-dot line represents $m_\Delta = 1.80 \, \text{GeV}$, and the dash-double-dot line depicts $m_\Delta = 1.88 \, \text{GeV}$. As expected, the results reveal an enhancement of the helicity amplitudes in the infrared region as the Delta mass increases, aligning more closely with the trend of the experimental data. However, the sensitivity to these mass variations remains relatively mild, with the helicity amplitudes converging to the original calculations at higher momentum transfers.


\section{SUMMARY}
\label{sec:Summary}

This work provides an in-depth presentation and discussion of the numerical results for the transition form factors and helicity amplitudes related to the process $\gamma^{(\ast)} + N(940)\frac{1}{2}^+ \to \Delta(1700)\frac{3}{2}^-$ by employing a symmetry-preserving treatment of a vector$\,\otimes\,$vector contact interaction (SCI) within the Dyson-Schwinger Equations (DSEs) framework, offering parameter-free results based on prior constraints. This approach models both baryons $N(940)\frac{1}{2}^+$ and $\Delta(1700)\frac{3}{2}^-$ as quark-diquark composites. By calculating transition form factors and helicity amplitudes, the study aims to provide insights into the internal structure of these baryons. The results of our analysis through the SCI provide a benchmark for more sophisticated QCD-based analyses which we plan to carry out in future.

Our key findings include the behavior of the magnetic dipole, $G_M^*$, electric quadrupole, $G_E^*$, and Coulomb quadrupole, $G_C^*$, transition form factors. At low and intermediate photon virtualities, the leading-order magnetic dipole form factor calculated within SCI-framework aligns well with experimental data. However, expected mild discrepancies appear at higher $Q^2$, where the experimental data suggest a sharper reduction of numerical results. It is in accordance with our expectation that the SCI computed form factors are harder than the QCD predictions.

While the electric and Coulomb quadrupole form factors qualitatively agree with experimental trends, significant deviations at low $Q^2$ suggest missing dynamics in the SCI framework, such as quark orbital angular momentum contributions due to Faddeev wave functions being independent of relative momenta, meson cloud effect, etc. The helicity amplitudes $A_{1/2}$, $A_{3/2}$, and $S_{1/2}$, derived from the form factors, further illustrate the SCI's performance. Overall, the study highlights the SCI framework's ability to reproduce key qualitative features of the transition, providing a foundation for more refined analyses based upon the running quark mass function.

Although the SCI model presents certain limitations, particularly in its exclusion of isovector-vector diquarks or orbital angular momentum components in the Faddeev wave function, the algebraic simplicity of the approach provides clear predictions that serve as useful benchmarks for future investigations that may include beyond-rainbow-ladder corrections and other frameworks built upon a Faddeev equation kernel and interaction vertices that possess QCD-like momentum dependence. The results presented herein are also expected to serve as a valuable reference for experimental facilities, such as the current JLab 12 GeV and its potential future 22 GeV upgrade. It can also spark exploration of studies involving higher resonance states within the broader context of QCD's non-perturbative regime.


\begin{acknowledgements}
L.A. acknowledges financial support provided by Ayuda B3 ``Ayudas para el desarrollo de l\'\i neas de investigaci\'on propias" del V Plan Propio de Investigaci\'on y Transferencia 2018-2020 de la Universidad Pablo de Olavide, de Sevilla.
G.P.T.  acknowledges financial support provided by the National Council for Humanities, Science and Technology (CONAHCyT), Mexico, through their program: {\em Beca de Posgrado en México}.
A.B. wishes to acknowledge the {\em Coordinaci\'on de la Investigaci\'on Cient\'ifica} of the {\em Universidad Michoacana de San Nicol\'as de Hidalgo}, Morelia, Mexico, grant no. 4.10, the {\em Consejo Nacional de Humanidades, Ciencias y Tecnolog\'ias}, Mexico, project CBF2023-2024-3544 as well as the Beatriz-Galindo support during his  scientific stay at the University of Huelva, Huelva, Spain. 
Otherwise, this work has been partially financed by 
Ministerio Español de Ciencia e Innovación under grant No. PID2022-140440NB-C22;
Junta de Andalucía under contract Nos. PAIDI FQM-370 and PCI+D+i under the title: “Tecnologías avanzadas para la exploración del universo y sus componentes" (Code AST22-0001).
\end{acknowledgements}


\bibliography{references}

\end{document}